\documentstyle{mn}
\newcommand{\acknowledgements}[1]{\vspace{7mm} \noindent {\normalsize \bf
Acknowledgements.\,} {\normalsize #1}}

\include{psfig} \psnoisy 

\begin{document}

\title[Galaxy evolution in the IR/submm] {Semi--analytic modelling of galaxy
evolution\\ in the IR/submm range}

\author[B. Guiderdoni et al.]  {Bruno Guiderdoni,$^1$ Eric Hivon,$^2$ Fran\c
cois R. Bouchet,$^1$ and Bruno Maffei$^3$ \\ $^{1}$ Institut d'Astrophysique
de Paris, CNRS, 98bis Boulevard Arago F--75014 Paris France\\ $^{2}$
Theoretical Astrophysics Center, Juliane Maries Vej 30, DK--2100 Copenhagen
Denmark\\ $^{3}$ Queen Mary and Westfield College, Mile End Road, London E1
4NS UK}

\maketitle

\begin{abstract}
This paper proposes a new semi--analytic modelling of galaxy properties in the
IR/submm wavelength range, which is explicitly set in a cosmological
framework. We start from a description of the non--dissipative and dissipative
collapses of primordial perturbations, and add star formation, stellar
evolution and feedback, as well as the absorption of starlight by dust and its
re--emission in the IR and submm. This type of approach has had some success
in reproducing the {\it optical} properties of galaxies. We hereafter propose
a simple extension to the IR/submm range. The growth of structures is followed
according to the standard Cold Dark Matter model. We assume
that star formation proceeds either in a ``quiescent'' mode, e.g. as in disks,
or in a ``burst'' mode with ten times shorter time scales. In order to
reproduce the current data on the evolution of the comoving cosmic SFR and gas
densities, we need to introduce a mass fraction involved in the ``burst'' mode
strongly increasing with redshift, probably reflecting the increase of
interaction and merging activity. We estimate the IR/submm luminosities of
these ``mild starburst'' and ``luminous UV/IR galaxies'', and we explore how
much star formation could be hidden in heavily--extinguished, ``ultraluminous
IR galaxies'' by designing a family of evolutionary scenarios which are
consistent with the current status of the ``cosmic constraints'',
as well as with the {\em IRAS} 60 $\mu$m luminosity function and 
faint counts, but
with different high--$z$ IR luminosity densities. However, these scenarios
generate a Cosmic Infrared Background whose spectrum falls within the $\pm
1\sigma$ range of the isotropic IR component detected by Puget {\it et al.}
(1996) and revisited by Guiderdoni {\it et al.} (1997). 
We give predictions for the faint galaxy counts and redshift
distributions at IR and submm wavelengths. The submm range is very sensitive
to the details of the evolutionary scenarios. As a result, the on--going and
forthcoming observations with {\em ISO} and SCUBA (and later with {\em SIRTF}, SOFIA,
{\em FIRST} and {\em PLANCK}) will put strong constraints on the evolution of 
galaxies at $z \sim 1$ and beyond.
\end{abstract}

\begin{keywords}
cosmology -- galaxy formation -- galaxy evolution -- infrared -- submm.
\end{keywords}

\section{Introduction}

This paper describes a new modelling of galaxy evolution in the infrared and
submm wavelength ranges, which are now open to high--redshift exploration by
the on--going observations with {\em ISO} and SCUBA, and forthcoming facilities and
experiments such as {\em SIRTF}, SOFIA, {\em FIRST} and {\em PLANCK}.
    
As a matter of fact, our knowledge of the early epochs of galaxies has
recently increased thanks to the richness and precision of the observational
evidence obtained by UV/visible/NIR surveys of high--redshift objects (Lilly
{\it et al.} 1995; Ellis {\it et al.} 1996; Cowie {\it et al.} 1996; Steidel
{\it et al.} 1996; Williams {\it et al.} 1996). The pattern of galaxy
evolution which emerges from these data can be summarized as follows: i) Faint
galaxy counts show the presence of a large number of blue objects, well in
excess of no--evolution expectations (Williams {\it et al.} 1996). ii) These
blue objects are ``sub--$L^\star$'' galaxies undergoing strong bursts of star
formation (Lilly {\it et al.} 1996; Ellis {\it et al.} 1996). iii) The fraction
of these blue objects with unclassified/peculiar morphologies showing signs of
tidal interaction and merging (Abraham {\it et al.} 1996) increases from local
samples to the HST high--resolution observations of the Medium Deep Survey
(Griffiths {\it et al.} 1994) and Hubble Deep Field (Williams {\it et al.}
1996). iv) The global star formation rate (hereafter SFR) of the universe
declined by a factor of about ten since redshift $z \sim 1$ (Lilly {\it et
al.} 1996; Madau {\it et al.} 1996; Sawicki {\it et al.} 1997; Connolly {\it
et al.} 1997).  v) Finally, this high SFR seems to be correlated to the
decrease of the cold--gas comoving density associated with damped
Lyman--$\alpha$ systems between $z=2$ and $z=0$ (Storrie--Lombardi {\it et
al.} 1996). These results nicely match a picture in which star formation in
bursts triggered by interaction/merging consumes the gas content of galaxies
as time goes on. It is common wisdom that such a qualitative scenario is
expected within the paradigm of hierarchical growth of structures. The
implementation of hierarchical galaxy formation in semi--analytic models
quantitatively substantiates this view and may further suggest that we have
seen the bulk of star formation and understood the broad features of galaxy
formation (Baugh {\it et al.} 1997).

However, it should be emphasized that this seemingly consistent view is
entirely based on visible/NIR observations which only probe the rest--frame
UV/visible light of high--$z$ objects. The total amount of energy released by
star formation should be estimated by summing up the UV/visible/NIR light of
stellar populations directly escaping from galaxies, {\em and} the part
which has been absorbed by dust and re--emitted in the IR/submm wavelength
range. The corrections needed to account for optical extinction by dust
are rather uncertain and could easily induce an upward revision of the
high--redshift SFR deduced from rest--frame UV/visible observations by factors
of a few.  Moreover, a significant fraction of star formation might be
completely hidden in heavily--extinguished galaxies which are missed by the
above--mentionned surveys.  Even for normal objects like spiral galaxies, the
fraction of energy emitted in the IR can amount to 30 \% of the optical, and
this can increase to more than 95 \% for ultraluminous starbursts. As a matter
of fact, the IR/submm range might be more adequate than the optical for
observing high--redshift, starburst galaxies provided they are dusty.  The
rest--frame IR/submm spectrum of dust heated by a young stellar population
peaks between 50 to 100 $\mu$m, pointing out the submm range as particularly
relevant for the search of primeval, high--redshift galaxies.

\begin{figure} 
\psfig{figure=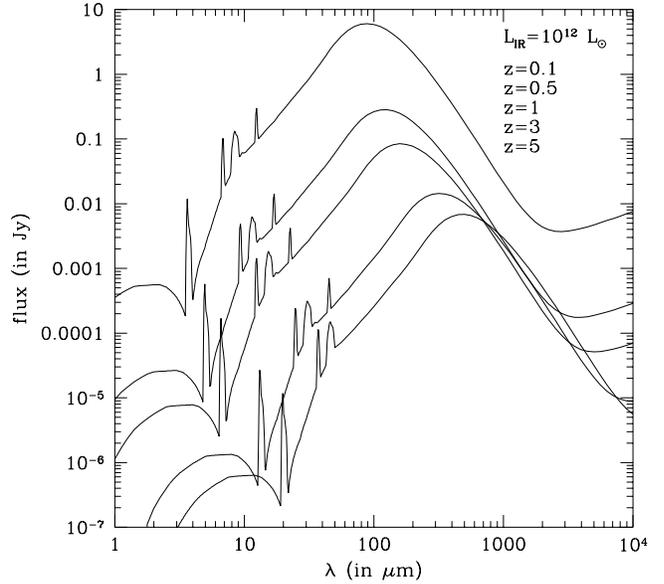,width=0.5\textwidth}
\caption{Observer--frame model spectra of a $L_{IR}=10^{12} L_{\odot}$ galaxy
at increasing redshifts (from top to bottom), for a cosmology
with $h=0.5$ and $\Omega_0=1$. The details of the modelling are
explained in Sect. 3.2. The reader is invited to note that the apparent flux
in the submm range is almost insensitive to redshift, because the shift of the
100 $\mu$m bump counterbalances the distance dimming.}
\end{figure}

At this time, we know very little about the ``optically--dark'' side of galaxy
evolution.  The {\em IRAS} satellite has given the first complete survey of 
FIR galaxy properties, in four bands between 12 and 100 $\mu$m. Many studies
have emphasized the wide variety of IR luminosities, from ``normal spirals''
to ``mild starbursts'', and finally to the ``luminous IR galaxies'' (hereafter
LIGs), mostly interacting systems, and the spectacular ``ultraluminous IR
galaxies'' (hereafter ULIGs), which are mergers (Sanders and Mirabel 1996;
Clements {\it et al.}  1996b). Unfortunately, while we have learned from {\em IRAS}
that about one third of the bolometric luminosity of the local universe is
released in the IR/submm (Soifer and Neugebauer 1991), we know very little
about galaxy evolution in this wavelength range. Faint galaxy counts and
redshift surveys down to flux densities $S_\nu \sim 60$ mJy (at 60 $\mu$m) do
not probe deeper than $z\sim 0.2$ (Ashby {\it et al.} 1996; Clements {\it et
al.} 1996a).  These surveys seem to show a strong luminosity and/or density
evolution of {\em IRAS} sources, but it is difficult to extrapolate this trend to
higher redshifts on a firm ground. In spite of its observational limits, {\em IRAS}
has also revealed the existence of {\em IRAS} 10214+4724,
a very peculiar, ``hyperluminous'' galaxy at $z=2.286$
(Rowan--Robinson {\it et al.} 1991a), though this object is
likely affected by lensing (Eisenhardt {\it et al.} 1996).  
It is expected that the {\em ISO} satellite will
complete and detail this picture, in a broader wavelength range from a few
$\mu$m to 200 $\mu$m. Other projects, such as {\em SIRTF}, will give access to
better sensitivity and imaging capabilities.

The properties of galaxies in the submm range are sensitive to the spectral
characteristics of dust, especially its emissivity at large wavelengths which
is not constrained by {\em IRAS} observations alone. With respect to
the relative wealth of data in the FIR, the submm emission of galaxies is
poorly known. The observational literature gathers the submm fluxes of only a
few tens galaxies which have been measured from ground--based or
aircraft--borne instruments. These observations are difficult and some of the
estimates of the amount of energy released in this range happen to be strongly
discrepant (e.g. Chini {\it et al.} 1986; Stark {\it et al.} 1989; Eales {\it
et al.} 1989; Chini and Kr\"ugel 1993). However, the observational 
situation will soon evolve with the start of SCUBA operations (and later,
with SOFIA and especially {\em FIRST}). About twenty counterparts of radiogalaxies 
and quasars have already been observed at submm/mm wavelengths 
(see e.g. Hughes {\it et al.}
1997; and references therein). Although their evolutionary status is not fully
understood, these objects give a first flavour of 
the future deep surveys which will be achieved by forthcoming instruments.

Theoretical modelling is needed to optimise observational strategies
and also to help assess the point source contribution of the forthcoming CMB
satellites MAP, and especially {\em PLANCK}. These experiments aim at mapping the
anisotropies of the submm/mm sky on scales well below the degree.  On these
scales, the separation of the various foregrounds and backgrounds which
superimpose to the fluctuations of the CMB is more difficult than for the
7$^\circ$ field of view of the very successful {\em COBE}/DMR experiment. In the
case of {\em PLANCK}, preliminary studies have shown 
that several thousands of galaxies
should be detected at several wavelengths between 400 $\mu$m and 1 cm.
  
\begin{figure} 
\psfig{figure=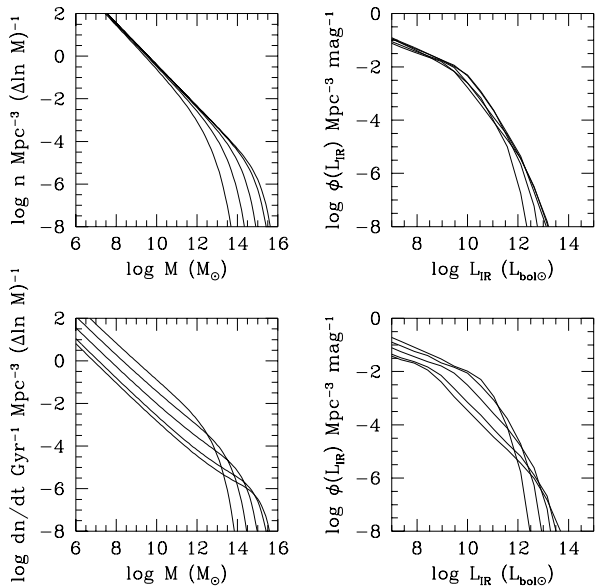,width=0.5\textwidth}
\caption{Number density and formation rate of collapsed haloes, and their 
relation to two modes of star formation. The
left--hand panels show the number density of collapsed haloes (top) and the
formation rate of collapsed haloes (bottom) computed for the SCDM model with
$h=0.5$, $\Omega_0=1$ and $\sigma_8=0.67$. The curves are plotted for
redshifts $z=3.67$, 1.99, 0.92, 0.23 and 0 (for increasing number densities at
the high--mass end). The right--hand top panel shows the evolution of the IR
luminosity function at the same redshifts, in the so--called ``quiescent'' mode of
star formation with $\beta=100$, $\alpha=5$ and $V_{hot}=130$ km s$^{-1}$.
The increase of the number of galaxies at all masses, as time goes on, is
similar to the increase of the number of haloes and reflects the accumulation
of galaxies. The right--hand bottom panel shows the evolution of the IR
luminosity function at the same redshifts, in the so--called ``burst'' mode of
star formation with $\beta=10$. The galaxies undergo strong starbursts which
act as beacons at the epochs of their initial collapses. Consequently, the
evolution of the luminosity function reflects the formation rate of new
haloes, with the characteristic crossing of the luminosity functions 
between low and high masses.}
\end{figure}

The epoch of galaxy formation can also be observed by its imprint on the
background radiation which is produced by the line--of--sight accumulation of
extragalactic sources. The search for the ``Cosmic Optical Background''
(hereafter COB) currently gives only upper limits. Nevertheless, the
shallowing of the faint counts obtained in the HDF suggests that we are now
close to convergence (Williams {\it et al.} 1996). Thus the lower limit to the
COB obtained by summing up the contributions of faint galaxies is likely
close to the real value. At longer wavelengths, the DIRBE instrument on {\em COBE}
has given upper limits on the FIR background between 2 and 300 $\mu$m (Hauser
1995), while Puget {\it et al.}  (1996) have discovered an isotropic component
in the {\em COBE}/FIRAS residuals between 200 $\mu$m and 2 mm, which
could be the
long--sought ``Cosmic Infrared Background'' (hereafter CIB). 
The presence of this component is confirmed by a new analysis restricted 
to the cleanest regions of the sky,
where the foreground Galactic components are negligible (Guiderdoni 
{\it et al.} 1997). Such a detection seems to yield the 
first ``post--{\em IRAS}'' constraint on the high--$z$ evolution of
galaxies in the IR/submm range, before the era of {\em ISO} results. Its level
comparable to the above--mentionned estimate of the COB suggests that a 
significant fraction of
the energy of young stars is absorbed by dust and released in the IR/submm.

The models which have been proposed to predict the faint counts and background
radiation in the IR/submm can be classified as ``backward evolution'' and
``forward evolution'', following the good review by Lonsdale (1996).  In the
first class of models, the IR/submm luminosity function undergoes luminosity
and/or number evolution which are simply parameterized as power laws of
$(1+z)$ (e.g. Weedman 1990; Beichman and Helou 1991; Hacking and Soifer 1991;
Oliver {\it et al.} 1992; Treyer and Silk 1993; Blain and Longair 1993;
Pearson and Rowan--Robinson 1996). These power laws are generally derived from
fits of {\em IRAS} galaxy counts (which do not probe deeper than $z \simeq
0.2$). Then they are extrapolated up to redshifts of a few
units. Unfortunately, various analyses of {\em IRAS} deep counts yield discrepant
results at $S_{60} < 300$ mJy, and the issue of the evolution is very
controverted (see e.g. Bertin {\it et al.} 1997, for a new analysis and
discussion).

\begin{figure} 
\psfig{figure=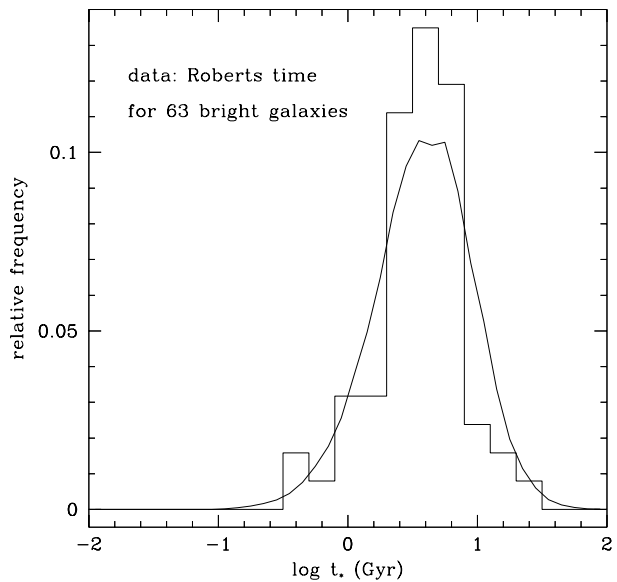,width=0.5\textwidth}
\caption{Distribution of the characteristic SFR time scales for the ``quiescent''
mode of star formation ($\beta=100$, $V_{hot}=130$ km s$^{-1}$ and $\alpha
=5$). Only galaxies with $V_c>V_{hot}$ are retained. The histogram shows the
data for a sample of 63 bright disk galaxies analysed by Kennicutt {\it et
al.} (1994).}
\end{figure}

In the second class of models, the photometric evolution in the IR/submm is
computed by implementing some of the involved physical processes.  For
instance: chemical evolution which rules the amount of dust responsible for
the IR/submm emission is modelled in Wang (1991a,b), and Eales and Edmunds
(1996a,b).  In addition to chemical evolution, the photometric evolution of
stellar populations which heat the dust is modelled by Franceschini {\it et
al.} (1991, 1994) and Fall, Charlot and Pei (1996). While the previous models
assume a simple relation between the dust content and the heavy--element
abundance of the gas, Dwek and Varosi (1996) try to explicitly model the
processes of dust formation and destruction.  However, both classes of models
assume that all galaxies form at the same redshift $z_{for}$ and that there is
no number evolution. But the paradigm of the hierarchical growth of structures
implies that there is no clear--cut redshift $z_{for}$ since galaxy formation
is a continuous process.  Only Blain and Longair (1993a,b) proposed a formalism
to compute the redshift range of galaxy formation, in addition to chemical
evolution,

A consistent approach to the early evolution of galaxies is particularly
important for any attempt at predicting their submm properties. Fig. 1 shows 
{\it model} spectra of a luminous IR galaxy as it would be observed
if placed at different
redshifts. There is a wavelength range, between $\sim 600$ $\mu$m and $\sim 4$
mm, in which the distance effect is counterbalanced by the ``negative
k--correction'' due to the huge rest--frame emission maximum at $\sim 100$
$\mu$m. In this range, the apparent flux of galaxies depends weakly on
redshift to the point that, evolution aside, a galaxy might be easier to
detect at $z=5$ than at $z=0.5$ ! The observer--frame submm fluxes, faint
galaxy counts and diffuse background of unresolved galaxies are consequently
very sensitive to the early stages of galaxy evolution. Note that this
particular wavelength range brackets the maximum of emission of the CMB.

\begin{figure} 
\psfig{figure=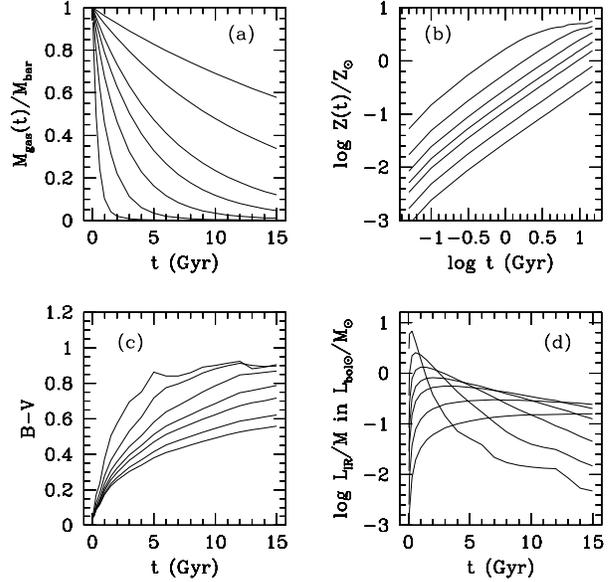,width=0.5\textwidth}
\caption{Evolution of various quantities for a grid of models with
$t_\star=0.33$, 1, 2, 3, 5, 10 and 20 Gyr.  Panel {\it a}: gas fraction (from
bottom to top).  Panel {\it b}: gas metallicity (from top to bottom).  Panel
{\it c}: $B-V$ colour (from top to bottom, taking into account face--on
internal extinction).  Panel {\it d}: luminosity re--emitted by dust in the
IR/submm (from top to bottom at 1 Gyr).}
\end{figure}

The dissipative and non--dissipative processes ruling galaxy formation in dark
matter haloes have been studied by various authors (see especially White and
Rees 1978; Schaeffer and Silk 1985; Evrard 1989; Cole 1991; Blanchard {\it et
al.} 1992). The modelling of these processes, complemented by star formation,
stellar evolution and stellar feedback to the interstellar medium, has been
achieved at various levels of complexity, in the so--called {\it
semi--analytic} approach which has been successfully applied to the prediction
of the statistical properties of galaxies (White and Frenk 1991; Lacey and
Silk 1991; Lacey {\it et al.} 1993, Kauffmann {\it et al.} 1993, 1994; Cole
{\it et al.} 1994; Heyl {\it et al.} 1995; Kauffmann 1995, 1996; Baugh {\it et
al.}  1996a,b, 1997). It turns out that, in spite of differences in the
details of the models, these studies lead to conclusions in the UV, visible
and (stellar) NIR which are remarkably similar.

None of these models has been applied so far to the prediction of the
properties of galaxies in the IR and submm ranges.  This is the main aim of
this paper which proposes a simple version of the semi--analytic approach and
gives predictions of faint counts at wavelengths between 60 $\mu$m and 1.4
mm. This first study also intends to identify some of the difficulties arising
in such a modelling. Sec. 2 quickly reviews the main physical processes which
have to be introduced in order to describe the formation of galaxies: the
non--dissipative collapse of perturbations, the dissipative collapse of the
gas, star formation and evolution, and feedback. Sec. 3 addresses the
peculiar issue of dust absorption and re--emission in the IR and submm.
Sec. 4 extracts useful information from the recent UV/visible deep surveys and
uses the new constraint arising from the CIB in order to generate a family of
evolutionary scenarios. Sec. 5 gives the IR/submm counts predicted from these
scenarios.  Finally, Sec. 6 briefly discusses these results, as well as the 
shortcomings of such models, and concludes.

In a previous study, we showed that the CIB can be disentangled with 
a family of evolutionary scenarios which predict steep submm counts, 
and that the observations with {\em ISO} (at 175 $\mu$m) will soon 
constrain the IR/submm evolution at $z \sim 1$ and beyond (Guiderdoni 
{\it et al.} 1997). Here we present the details of the modelling and 
apply it to other observations, more specifically in SCUBA bands 
(through the narrow submm atmospheric windows).
A preliminary version of this work was presented in Guiderdoni {\it et al.}
(1996).
A forthcoming paper (Hivon {\it et al.} 1997) will show simulations of
the anisotropies of the diffuse submm background due to galaxies and will
study the possibility of their detection, with current and forthcoming
instruments.  Other papers in this series will try to overcome some of the
shortcomings of this first study.

\begin{figure}
\psfig{figure=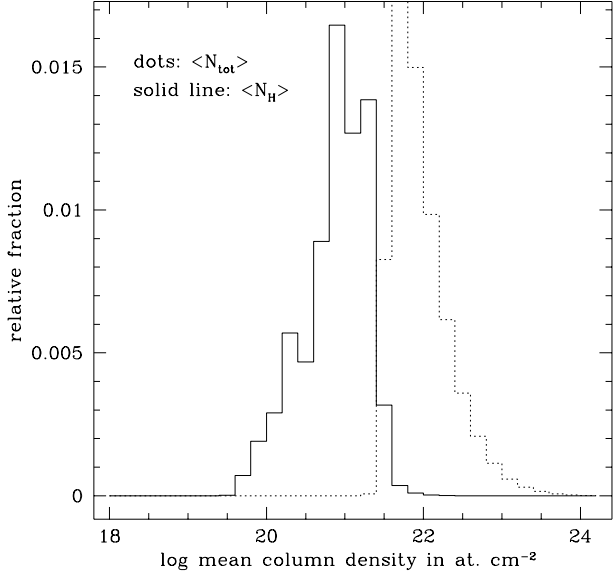,width=0.5\textwidth}
\caption{Distribution of mean column densities at $z=0$, in atom cm$^{-2}$,
for galaxies with $SB_e < 25$ mag arcsec$^{-2}$, and the ``quiescent'' mode of
star formation ($\beta=100$, $\alpha =5$ and $V_{hot}=130$ km s$^{-1}$).
Dotted line: $\log <N_{tot}>$; solid line: $\log <N_{H}>$.}
\end{figure}

\begin{figure}
\psfig{figure=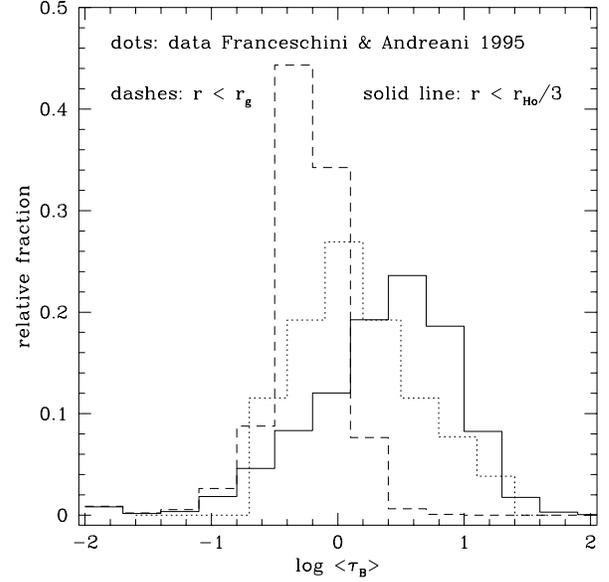,width=0.5\textwidth}
\caption{Distribution of mean, face--on optical depth in the $B$--band at
$z=0$, for the ``quiescent'' mode of star formation ($\beta=100$, $\alpha =5$ and
$V_{hot}=130$ km s$^{-1}$).  Two ways of computing $\tau_B$ are
considered. Dashes: Mean optical depth inside the gas radius
($r_g=1.6r_{25}$). Solid line: Mean optical depth inside one third of the
Holmberg radius (at 26.5 mag arcsec$^{-2}$). The dotted line shows the
observational distribution in a sample of normal spirals and
luminous IR galaxies from
Franceschini and Andreani (1995). The total dust quantity derived from 1.3 mm
observations is distributed within one third of the Holmberg radius and an
observational estimate of $\tau_B$ is computed. The observational distribution
peaks at the value $<\tau_B> \sim 1$ and is bracketted by the two predicted
histograms.}
\end{figure}

\section{A Schematic view of galaxy formation}

\subsection{Non--dissipative collapse}

The formation and evolution of a galaxy in its dark matter halo can be briefly
sketched as follows: the initial perturbation, which is gravitationally
dominated by non--baryonic dark matter, grows and collapses. After the
(non-dissipative) collapse, and subsequent violent relaxation, the halo
virializes, through the formation of a mean potential well seen by all
particles, which consequently share the same velocity distribution. As
detailed in the Appendix, we use the classical top--hat model for
spherically--symmetric perturbations, and we compute the mass distribution of
collapsed haloes from the peaks formalism (Bardeen {\it et al.} 1986), as in
Lacey and Silk (1991) and Lacey {\it et al.} (1993).  Hereafter, we shall
consider the SCDM model with $H_0=50$ km s$^{-1}$ Mpc$^{-1}$, $\Omega_0=1$,
$\Lambda=0$, and $\sigma_8=0.67$ as an illustrative case. We take a baryonic
fraction $\Omega_{bar}=0.05$, consistent with primordial nucleosynthesis. The
redshift evolution of the number density and formation rate of collapsed
haloes is plotted in fig. 2.

\subsection{Cooling and dissipative collapse}

The baryonic gas cools in the potential wells of dark matter haloes, by a
process identical to cooling flows observed at the centre of rich clusters.
The cooling time at halo radius $r$ is
\begin{equation}
t_{cool}(r) = {3 \over 2} {n_{tot}(r) kT_V \over n_e^2(r) \Lambda(T_V)} = 3
\pi G \mu_e^2 m_p^2 {\Omega_0 \over \Omega_{bar}} {r^2 \over \Lambda (T_V)} ,
\end{equation}
with $\mu_e=1.14$ for ionized primordial gas. The cooling curve $\Lambda (T)$
takes into account various cooling processes. We do not consider the
metallicity dependence of $\Lambda (T)$. Neglecting this dependence leads to
an {\it overestimate} of cooling times, which are already very short.  The
equation $t_{cool}(r_{cool}) = t(z)$ defines a cooling radius $r_{cool}$ as a
function of redshift $z$. At this redshift, only gas inside $r_{cool}$ (or
$r_V$ if $r_{cool} > r_V$) cools and is available for star formation.  This
cooling criterion introduces a high--mass cut-off in the mass distribution of
cold baryonic cores. At the low--mass end, the cooling is so efficient that
almost all the gas can cool, leading to a slope of the mass distribution of
baryonic cores $n(M_{bar})dM_{bar} \propto M_{bar}^{-1.95}dM_{bar}$ at
redshift $z=0$, which is close to the slope $n(M)dM \propto M^{-2}dM$ for the
number density of collapsed haloes. This is the so--called ``overcooling''
problem (Cole 1991; Blanchard {\it et al.} 1992).

The final radius of the cold gas in rotational equilibrium is related to the
initial radius by conservation of angular momentum (Fall and Efstathiou 1980).
Approximately, $r_D \sim \lambda \min (r_V, r_{cool})$, with the dimensionless
spin parameter $\lambda \equiv J|E|^{1/2}G^{-1}M^{-5/2}\simeq 0.05 \pm 0.03$
(Barnes \& Efstathiou 1987; Efstathiou {\it et al.} 1988; Zurek {\it et al.}
1988). Previous studies only used the mean value of $\lambda$ in this formula.
Hereafter, we introduce the $\lambda$ distribution from Barnes and Efstathiou
(1987) model C0--4 (their fig. 11).  According to a fit based on fig. 3 of
Fall and Efstathiou (1980), the exponential disk which forms from the
dissipative collapse of the gas has a length scale $r_0 \simeq 1.26
\lambda^{1.17}\min (r_V, r_{cool})$ and a radius including half the cold
baryonic mass $r_{1/2}/r_0=1.68$.  A dynamical time scale in the disk--like
core is $t_{dyn} \equiv 2\pi r_{1/2} /V_c$.  It is important to note that only
disks can form in this formalism. The formation of elliptical galaxies (and of
bulges of spiral galaxies) has to be explained by the merging of
disks. Kauffmann {\it et al.} (1993) and Cole {\it et al.} (1994) showed that
this merging process can easily explain the current fraction of gE among
bright galaxies.
 
\subsection{Star formation}

Locally, the SFR depends on numerous physical parameters.  Nevertheless,
phenomenological studies seem to show that, on galaxy scales, the SFR per unit
surface density is proportional to the total gas surface density (neutral plus
molecular) divided by the dynamical time scale of the disk (Kennicutt 1989,
1997). So we shall hereafter assume that the star formation time scale
$t_\star$ is proportional to the dynamical time scale of the disk $t_{dyn}$
and we introduce a first efficiency factor $\beta$. With $t_\star \equiv \beta
t_{dyn}$, we take:
\begin{equation}
SFR(t) = {M_{gas}(t) \over t_\star} .
\end{equation}
Fig. 3 shows the predicted $t_\star$ histogram compared to the histogram of
``Roberts times'' for a sample of 63 bright disk galaxies observed by
Kennicutt {\it et al.} (1994). The Roberts time is defined as $t_R \equiv
(M_{HI}+M_{H_2})/SFR(t_0)$ where $M_{HI}$ is the gas mass in the $HI$ phase
measured from the 21 cm line, $M_{H_2}$ is the gas mass in the $H_2$ phase
measured from the CO line, and $SFR(t_0)$ is the total star formation rate
measured from the H$\alpha$ line, under an assumption about the shape of the
Initial Mass Function (hereafter this is Salpeter IMF).  The interesting
result is that the model correctly predicts the {\it shape} and {\it width} of
the histogram.  We emphasize that this agreement is due to both the range of
halo densities (scaling as $(1+z_{coll})^3$) {\it and} dimensionless spin
parameters $\lambda$.  Without the $\lambda$ scatter, the predicted
distribution would be about three times too narrow. The average value of the
observed histogram can be reproduced by taking $\beta \simeq 100$ for our SCDM
model.

Finally, for sake of simplicity, we use a Salpeter IMF with index $x=1.35$.
Stars have masses $0.1 \leq m \leq 120 M_\odot$. We also assume that the mass
fraction blocked in dark objects with masses below 0.1 $M_\odot$ is
negligible.

\subsection{Stellar feedback}

The explosion of massive stars can expel gas from the galaxies and quench star
formation, leading to a strong increase of the mass--to--luminosity ratios in
small objects.  Observationally, HI holes and X--ray superbubbles are good
evidence that such galactic winds are present in galaxies. The stellar
feedback is introduced in a similar way by most authors, following the
original work by Dekel and Silk (1986). By equating the gas binding energy to
the thermal energy ejected by supernovae, one gets:
\begin{equation}
{1 \over 2}M_{gas}(t) ({V_{esc} \over V_c})^2 V_c^2 = \epsilon \int_0^{t_W}
\tau_*(t') \eta_{SN} E_{SN} dt' ,
\end{equation}
where $\eta_{SN}$ is the number of SNe per unit mass of stars, depending on
the IMF. For our choice of IMF, $\eta_{SN}=7.4~10^{-3}$ $M_\odot^{-1}$. The
output mechanical energy of a SN is $E_{SN} \sim 10^{51}$ erg.  The escape
velocity at radius $r \leq r_V$ in a singular isothermal sphere truncated at
radius $r_V$ is $V_{esc}(r)=\sqrt 2 V_c (1-\ln (r/r_V))^{1/2}$.  The maximum
of the $\lambda$ distribution corresponds to $r_{1/2}/r_V \simeq 0.05$,
leading to $V_{esc}/V_c \simeq 2.8$.  Since much of this energy is
subsequently radiated away, we insert a second efficiency factor $0 \leq
\epsilon \leq 1$.  After some algebra, the mass fraction of stars $F_\star$
forming before the triggering of
the galactic wind at time $t_W$ is given by:
\begin{equation}
F_\star \equiv {M_\star(t_W) \over M_\star(t_W) + M_{gas}(t_W)} 
= (1 + (V_{hot}/V_c)^\alpha)^{-1} ,
\end{equation}
with $\alpha =2$ and $V_{hot}\equiv
(V_c/V_{esc})(2\eta_{SN}E_{SN})^{1/2}\epsilon^{1/2} = 310\, \epsilon^{1/2}$ km
s$^{-1}$ for the Salpeter IMF.  Nevertheless, there is much uncertainty on
these parameters because of the cooling of supernova remnants before wind
triggering. For instance, numerical simulations seem to suggest that $\epsilon
\sim 0.1$ (Thornton {\it et al.} 1997). Cole {\it et al.} (1994) introduced a
fit based on SPH simulations of galaxy formation in which most of the feedback
effect is due to momentum exchange rather than to ISM heating (Navarro and
White 1993). For a typical feedback parameter $f_V=0.1$, the numerical
simulations can be fitted with $\alpha =5$ and $V_{hot}=130$ km s$^{-1}$. We
shall hereafter take these values as our standard parameters. The situation is
still complicated by the existence of {\it non--local} feedback processes in
addition to the local ones. Efstathiou (1992) and Blanchard {\it et al.}
(1992) suggest that high--$z$ re--ionization of the IGM could prevent
cooling in haloes with circular velocities below $V_{equ} \sim 20$ to 50 km
s$^{-1}$, and possibly as high as $\sim (200)^{1/3} V_{equ}$ in case of
adiabatic collapse. So it is very likely that the overall quenching of dwarf
formation depends on redshift. Introducing this redshift dependence of dwarf
formation partly alleviates the problem of the steep slope of the luminosity
function (see e.g.  Kauffmann {\it et al.} 1993). So the situation appears to
be very complicated, in the absence of a global theory of feedback processes.
In the following, we shall simply model the combination of
local and non--local feedbacks by
introducing a simple $(1+z_{coll})$ dependence for $V_{hot}$.

\begin{figure}
\psfig{figure=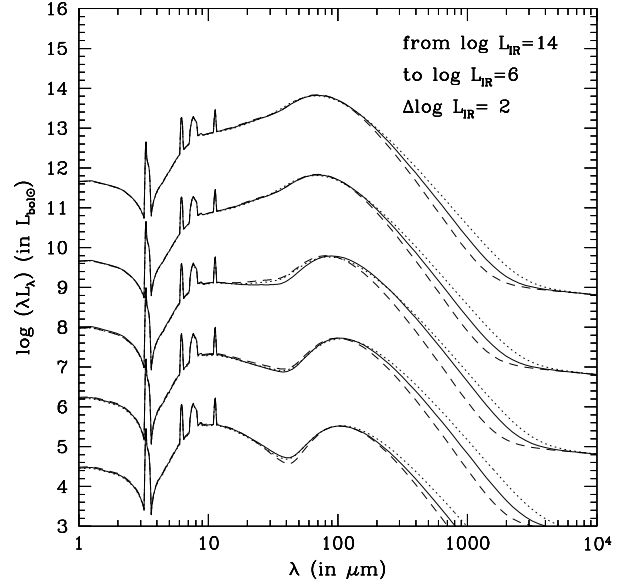,width=0.5\textwidth}
\caption{Model spectra in the IR and submm, for IR luminosities $10^6$,
$10^8$, $10^{10}$, $10^{12}$, and $10^{14} L_{bol\odot}$.  Emissivity index of
big grains: $m=2$ (dashes), $m=1.5$ (solid lines), $m=1$ (dotted lines).}
\end{figure}

\begin{figure} 
\psfig{figure=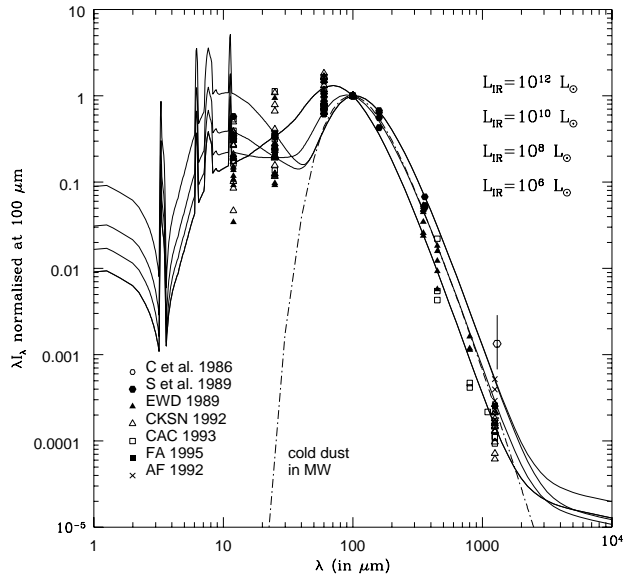,width=0.5\textwidth}
\caption{Model spectra with $m=1.5$ superimposed to a compilation of data in
the FIR and submm: Chini {\it et al.} 1986 (mean value), Stark {\it et al.}
1989, Eales {\it et al.}  1989, Carico {\it et al.} 1992, Andreani and
Franceschini 1992, Clements {\it et al.} 1993, Franceschini and Andreani 1995
(mean value). The dashes--and--dots show a typical spectrum of cold dust in
the Milky Way (Reach {\it et al.}  1995). Models and data are normalised at
100 $\mu$m.}
\end{figure}

\subsection{Spectral evolution of the stellar population}

A model of spectrophotometric evolution is used to compute the age dependence
of the gas content, the UV to NIR spectra of the stellar populations, and the
mass--to--luminosity ratios.  The stars are placed on the Zero--Age Main
Sequence of the HR diagram according to the IMF. The models use compilations
of stellar evolutionary tracks taking into account the various stages of
stellar evolution in order to compute at each time step the distribution of
stellar populations in the HR diagram. This distribution is combined with
a library of stellar spectra and gives the synthetic spectrum
$F_{\star\lambda}$. At the end of their lifetimes, stars die and return a
fraction of their mass to the ISM. The model which is used here is described
in Guiderdoni and Rocca--Volmerange (1987, 1988), and Rocca--Volmerange and
Guiderdoni (1988), and includes upgraded stellar tracks from Schaller {\it et
al.} (1992) and Charbonnel {\it et al.} (1996).

Panel {\it a} of fig. 4 shows the grid of models with various $t_\star$ from
0.33 to 20 Gyr, which are introduced in order to follow the relative gas
content of the galaxy $g(t) \equiv M_{gas}(t)/M_{bar}$. The heavy elements
synthetised by stars are injected into the interstellar medium.
The metallicity of the
gas is estimated from the Instant Recycling Approximation $Z_g(t)=-y_Z \ln
g(t)$ with a yield $y_Z$ depending on the choice of the IMF. Panel {\it b}
shows the time variation of $Z_g(t)$ for the grid of models.  The photometric
properties can be computed after taking into account the intrinsic extinction
(see the following section). As an example, panel {\it c} gives the face--on
$B-V$ colours.

\subsection{The slope of the luminosity function}

Before examining the modelling of the IR emission of galaxies, we would like
to quickly comment on the slopes of the $B$--band luminosity and gas mass
functions obtained by this simple version of the semi--analytic approach.

In spite of the variety of SFR histories, and of the strong variation of the
mass--to--luminosity ratios in time scales of a few Gyr, the {\it shape} of
the luminosity function in the absence of feedback processes is 
surprisingly similar to that of the baryonic mass function. 
For $\phi (L_B)dL_B \propto L_B^{s} dL_B$, our model with $\beta=100$ and the 
standard mass loss ($\alpha=5$ and $V_{hot}=130$ km s$^{-1}$) gives
$s=-1.4$, whereas Loveday {\it et al.} (1992) find $s=-1$.
Such an uncomfortable
situation is a robust result of {\it all} the recent attempts to model galaxy
formation and evolution in a semi--analytic approach (see Kauffmann {\it et
al.} 1994; Cole {\it et al.} 1994). Although it seems in disagreement with the
nearby surveys (but other surveys seem to suggest an increase of the slope at
the faint end, e.g. Marzke {\it et al.} 1994), this steep slope is necessary
to reproduce the deep redshift surveys (down to $B_J < 24$) and the faint
galaxy counts (down to $B_J < 28$).  Subtle selection effects due to surface
brightness could explain the discrepancy between the nearby luminosity
function and the high--$z$ one (McGaugh 1994; Lobo and Guiderdoni 1997).

Indeed, this type of selection effect can be suspected because the predicted
gas mass function is in better agreement with the observational data.  The
predicted mass distribution of gas at redshift $z=0$ has a slope
$n(M_{gas})dM_{gas} \propto M_{gas}^{-1.3}dM_{gas}$ (with the standard values
or the parameters). Smaller objects form earlier on an
average, with higher densities, smaller $t_{dyn}$, smaller $t_\star$, and, 
consequently, they are more affected by mass loss and their relative gas 
content is lower. It is possible to correct
statistically the total gas mass function in order to predict an HI mass
function and compare it to the observational slope $-1.35$ determined
by Briggs and Rao (1993). 
In the sample of 63 disk galaxies gathered by Kennicutt {\it et al.}
(1994), there is no systematic trend for $M_{HI}/M_{gas}$ versus $M_{gas}$,
$t_R$, or the $SFR$. We have made an histogram of $\log (M_{HI}/M_{gas})$,
with values from $-1$ to 0, and we have distributed the predicted number
density at $M_{gas}$ into the range of $M_{HI}$ values according to this
histogram. The effect of this correction turns out to be rather small.  Thus
there is a good agreement of the predictions with the observations. It is
worthwhile to emphasize that the slope $-1.35$ of the data depends on the
volume corrections of the survey for low--mass objects and is somewhat
uncertain.

\begin{table*}
\caption{Scenarios of galaxy evolution}
\begin{center}
\begin{tabular}{@{}lrrrl} \hline
Name & $f_{burst}$ & $f_{quiescent}$ & \% of ULIGs & Line code\\ 
& ($\beta=10$)&($\beta=100$)& &\\ 

Q & 0 & 1 & 0 \% & dots and small dashes\\ 
A & $0.04(1+z_{coll})^5$ & $1-f_{burst}$ & 0 \% & solid line\\ 
B & $0.04(1+z_{coll})^5$ & $1-f_{burst}$ & 5 \% at all $z_{coll}$ & dotted line\\ 
C & $0.04(1+z_{coll})^5$ & $1-f_{burst}$ & 90 \% for $z_{coll} > 3.5$ & long dashes\\ 
D & $0.04(1+z_{coll})^5$ & $1-f_{burst}$ & 15 \% at all $z_{coll}$ & short dashes\\ 
E & $0.04(1+z_{coll})^5$ & $1-f_{burst}$ & $1- \exp -0.02(1+z_{coll})^2$ & dots and long dashes\\
\end{tabular}
\label{ta:mod}
\end{center}
\end{table*}

\section{Spectral evolution of dust emission}

\subsection{Dust absorption}

Part of the energy released by stars is absorbed by dust and re--emitted in
the IR and submm ranges.  The derivation of the IR/submm spectrum is a
three--step process: i) computation of the optical thickness of the disks; ii)
computation of the amount of bolometric energy absorbed by dust; iii)
computation of the spectral energy distribution of dust emission.  The
modelling of these steps is not an easy task since it requires addressing
confused issues such as the chemical evolution of the dust, and the
geometrical distribution of dust relatively to stars.

We assume that the gas is distributed in an exponential disk with truncation
radius $r_g$ and mean $H$ column density $<N_H(t)>=M_{gas}(t)/1.4 m_H \pi
r_g^2$.  The factor 1.4 accounts for the presence of helium.  If $r_{25}$ is
the isophotal radius at 25 mag arcsec$^{-2}$, the observations give
$r_g/r_{25}\simeq 1.6$ (Bosma 1981). The $r_{25}$ radii are consistently
computed from the $B$ magnitudes and $r_0$ radii of the galaxies. Fig. 5 shows
the mean total and gas surface densities inside $r_g$ at redshift $z=0$.
These surface densities fairly correspond to the crude estimate
$<N_H(t)>=6.8~10^{21} g(t)$ atom cm$^{-2}$ used in Guiderdoni and
Rocca--Volmerange (1987) and Franceschini {\it et al.} (1991, 1994).  As noted
by Guiderdoni \& Rocca--Volmerange (1987), a galaxy with $g \simeq 0.20$ has
an H column density $<N_H> \simeq 1.4~10^{21}$ atom cm$^{-2}$ in good
agreement with the observational value for late--type disks, in spite of the
uncertainties in this estimate (Guiderdoni 1987).

The mean optical thickness inside $r_g$ is given by:

\begin{eqnarray}
\tau_\lambda (t) &=& {1 \over 1.086} {A_\lambda \over A_V} (Z_{g}(t)) {A_V
\over E_{B-V}} {E_{B-V} \over N_H} <N_H(t)> \\ &=& ({A_\lambda \over
A_V})_{Z_\odot} ({Z_g(t) \over Z_\odot})^s ({<N_H(t)> \over
2.1~10^{21}~at~cm^{-2}}) .
\end{eqnarray}

As in Guiderdoni and Rocca--Volmerange (1987) and Franceschini {\it et al.}
(1991, 1994), the extinction curve depends on the gas metallicity $Z_g(t)$
according to power--law interpolations based on the Solar Neighbourhood and
the Magellanic Clouds, with $s=1.35$ for $\lambda < 2000$ \AA\ and $s=1.6$ for
$\lambda > 2000$ \AA.  The extinction curve for solar metallicity is taken
from Mathis {\it et al.} (1983).

It is not clear whether the disks of ``normal'' spirals are optically thin or
optically thick. Catalogues of galaxies (such as the classical RC2 and the
RSA) suggest $\tau_B =0.7$ for spirals.  On the other hand, several studies
have pointed out the possibility of optically--thick disks (see e.g. Disney
{\it et al.} 1989 and references therein). Since the gas decrease is
counterbalanced by the metallicity increase, the optical thickness of
our grid of models is maximal
for $g(t)=e^{-s} \simeq 0.20$ with $s=1.6$. Late--type disks spend most of
their life time at optical depths $\tau_B \simeq 0.4$ -- 0.7, in good
agreement with what is suggested in the RC2 or RSA, and with the estimates of
Rowan--Robinson (1992) based on {\em IRAS} results.  Fig. 6 gives the predicted
distribution of optical thicknesses which fairly compares with a sample of
normal spirals and luminous IR galaxies (Franceschini and Andreani 1995).

As in Guiderdoni and Rocca--Volmerange (1987) and Franceschini {\it et al.}
(1991, 1994), we also assume a simple geometric distribution where the gas and
the stars which contribute mainly to dust heating are distributed with equal
height scales in the disks.  If $\tau_\lambda (t)$ is the optical thickness of
the disks at wavelength $\lambda$ and time $t$, the extinction correction
(averaged over inclination angle $i$) is:
\begin{equation}
A_\lambda (t) = -2.5 \log < {1 - \exp (-a_\lambda \tau_\lambda (t) /\cos i)
\over a_\lambda \tau_\lambda (t) /\cos i} >_i ,
\end{equation}
providing that the stars and gas have the same height scales.  The factor
$a_\lambda \equiv (1-\omega_\lambda)^{1/2}$ crudely takes into account the
effect of the albedo $\omega_\lambda$ (from Draine and Lee 1984)
for isotropic scattering (Natta and
Panagia 1984) and relates extinction to absorption.  This ``slab'' geometry is
intermediate between the ``screen'' geometry $\exp -a_\lambda \tau_\lambda
/\cos i$ and the ``sandwich'' geometry (with zero height scale for dust)
$(1+\exp -a_\lambda \tau_\lambda /\cos i)/2$ which respectively lead to larger
and smaller absorption. Observational analyses seem to suggest that this
``slab'' geometry is the best guess to fit the data (see e.g. Franceschini and
Andreani 1995; Andreani and Franceschini 1996).

Finally, the bolometric luminosity which is absorbed by dust and released in
the IR/submm is:
\begin{equation}
L_{IR}(t) = \int F_{\star \lambda} (t) (1-10^{-0.4A_\lambda(t)})d\lambda .
\end{equation}
Panel {\it c} and {\it d} of fig. 4 respectively show the face--on 
extinguished
$B-V$ colours and the IR bolometric luminosities per unit mass of galaxy for a
grid of models with various $t_\star$. The corresponding range of IR--to--blue
luminosity ratios $0.06 \leq L_{IR}/\lambda_B L_B \leq 4$ is very similar to
the observed range for optically--selected samples, for instance the sample
drawn from the Zwicky Catalogue with $m_z<$ 15.7 mag in Soifer {\it et al.}
(1987), and characterizes ``mild starbursts'' and LIGs.  The IR--to--blue
luminosity ratios are sensitive to $\tau_B$ and the geometry. If
there are tidally--induced gas inflows resulting in smaller $r_g$ and to a
``screen''--like gas distribution, this ratio can easily reach 10--100 as in
the ULIGs. We hereafter keep the LIG--type efficiency of conversion between
the optical and the IR as a conservative estimate, and we shall consider the
population of ULIGs in Sec. 4.3.

\begin{figure} 
\psfig{figure=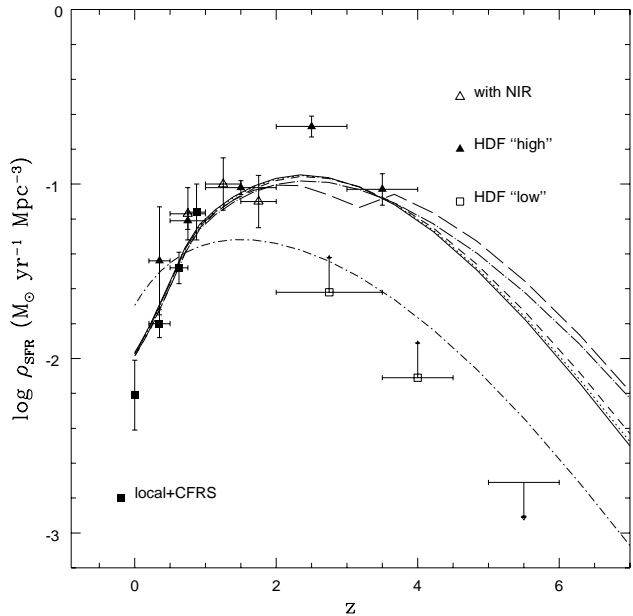,width=0.5\textwidth}
\caption{Evolution of the cosmic comoving star formation rate density as
computed from rest--frame UV luminosity densities, by using Salpeter IMF with
slope $1.35$. The UV densities are based on the integration of the 
luminosity function fit on all magnitudes.
The UV fluxes have {\it not} been corrected for intrinsic extinction. 
The solid squares are derived from the rest--frame 2800 \AA\
luminosity density in the Canada--France Redshift Survey, and the local value
is from the $B$--band luminosity function and average UV-$B$ colour (Lilly
{\it et al.} 1996). The open triangles take into account NIR data to compute
photometric redshifts in the Hubble Deep Field (Connolly {\it et al.}
1997). The solid triangles come from another photometric--redshift analysis of
the HDF (Sawicki {\it et al.} 1997), while the open squares
with the arrows are lower
and upper values from the HDF with redshifts derived from Lyman--continuum
drop--outs (Madau {\it et al.} 1996).  Scenario Q (``quiescent'' mode,
$\beta=100$) is plotted with dots and small dashes. It is unable to reproduce
the steep decline of $\rho_{SFR}$ between redshifts 1 and 0. Other scenarios
involve an increasing fraction of the ``burst'' mode ($\beta=10$) with
redshift, producing a population of LIGs.  Scenario A (solid line) has no
ULIGs. Various quantities of ULIGs are included in scenarios B (dotted line),
C (long dashes), D (short dashes) and E (dots and long dashes). 
See tab. 1 for details and summary of lines codes. The SFRs slightly differ 
because $t_\star$ is considered as an
exponential time scale for the LIGs and a burst duration for the ULIGs.}
\end{figure}

\subsection{Dust emission}

The emission spectra of galaxies are computed as a sum of various components,
according to the method developped by Maffei (1994), which uses the
observational correlations of the {\em IRAS} flux ratios 12$\mu$m/60$\mu$m,
25$\mu$m/60$\mu$m and 100$\mu$m/60$\mu$m with $L_{IR}$ (Soifer and Neugebauer
1991).  These correlations are extended to low $L_{IR}$ with the samples of
Smith {\it et al.} (1987) and especially Rice {\it et al.} (1988).

Several components are considered in the model spectra:

\begin{itemize}

\item Polycyclic aromatic hydrocarbons (PAH). Because of their small size
($\leq $ 1 nm), these molecules never reach thermal equilibrium when they are
excited in a UV/visible radiation field. Their temperature fluctuates and can
reach a value much higher than the equilibrium temperature, explaining the 12
$\mu$m excess and the bands at 3.3, 6.2, 7.7, 8.6 and 11.3 $\mu$m.  Their
template emission spectrum is taken from the model by D\'esert {\it et al.}
(1990).
\item Very small grains (VSG). They are made of graphite and silicates. These
dust grains have sizes between 1 and 10 nm. As the PAH, they never reach
thermal equilibrium. Consequently, their emission spectrum is much broader
than a modified black body spectrum at a single equilibrium temperature. The
template emission spectrum is also taken from the model by D\'esert {\it et
al.} (1990).
\item Big grains (BG). They are also made of graphite and silicates. These
dust grains have sizes between 10 nm and 0.1 $\mu$m.  They (almost) reach
thermal equilibrium and can be reasonably described by a modified black body
$\epsilon_\nu B_\nu(T_{BG})$ and emissivity $\epsilon_\nu \propto \nu^m$ with
$1 \leq m \leq 2$.
\item Synchrotron radiation. This non--thermal emission is strongly correlated
with stellar activity and, as a consequence, with IR luminosity (see
e.g. Helou {\it et al.} 1985).  According to observations at 1.4 GHz, this
correlation is $L_\nu(1.4~GHz)=L_{IR}/(3.75~10^{12} \times 10^q)$. Here
$L_{IR}$ is in W, $\nu_{80}=3.75~10^{12}$ Hz is the frequency at 80 $\mu$m and
$q\simeq 2.16$ is determined from observations. Then we assume that we can
extrapolate from 21 cm down to $\sim $ 1 mm with a single average slope 0.7,
so that $L_\nu=L_\nu(1.4~GHz)(\nu /1.4~GHz)^{-0.7}$.
 
\end{itemize}

The $60/100$ colour gives the temperature $T_{BG}$ of the BG, provided that an
emissivity index $m$ has been chosen. We hereafter test $m=1$, 1.5 (standard
value) and 2. Then the amount of BG, VSG and PAH are calculated iteratively
from the $12/100$, $25/100$ and $60/100$ ratios. The resulting spectra are
computed from a few $\mu$m to several mm and evolve with $L_{IR}$ such as more
luminous galaxies preferentially emit at shorter wavelengths.  By
construction, these spectra fit the {\em IRAS} colour correlations.  
Fig. 7 shows
examples of these model spectra for various $L_{IR}$ and three values of the
emissivity index for BG.

In spite of its shortcomings, this method takes into account the observed
spectral evolution in the IR and submm/mm ranges, while previous works
generally use a single template spectrum in the whole spectral range.  It is
also possible to reproduce the FIR photometry of various individual galaxies
by this method (Maffei 1994). Fig. 8 shows a compilation of data superimposed
to a series of model spectra.  The compiled samples gather galaxies which have
been observed at least at one wavelength in the submm range, (say, between 200
$\mu$m and 1.3 mm) with a positive detection. The various samples are very
heterogeneous. They have been obtained with different telescopes and
instruments and focus upon different types of galaxies (``normal'' spirals,
LIGs, ULIGs). None of them is complete.  Moreover, there is very little overlap
between the samples, and the few galaxies which have been observed by various
authors with various instruments generally have discrepant fluxes. These
discrepancies can originate from the choices of beam size and beam shopping.
Andreani and Franceschini (1992, 1996) have studied the effect of the finite
extension of the submm emission and computed an average aperture correction,
which turns out to be rather small.

There is some debate about the presence of very cold dust which could emit at
submm wavelengths. Chini {\it et al.} (1986), confirmed by Chini and Kr\" ugel
(1993), claim that the dust seen at {\em IRAS} wavelengths is unable to reproduce
the observed submm emission.  On the contrary, Stark {\it et al.} (1989),
Eales {\it et al.} (1989), and Clements {\it et al.} (1993) do not detect any
evidence for this new component.  Chini {\it et al.} (1986) interpret their
high fluxes as an evidence for the existence of cold dust. The value of the
derived temperature depends on the choice of the emissivity index of these
grains, which can be anything between 1 and 2.  Clearly, the uncertainties in
the submm emission are large, and the need for a systematic survey of a
complete sample is strong. There is no doubt that this will be one of the
first targets of SCUBA. Because of these uncertainties, the models with an
average index $m=1.5$ plotted in fig. 8 seem to cover the range of
observations at submm wavelengths, except the high fluxes observed by Chini
{\it et al.} (1986).

\begin{figure} 
\psfig{figure=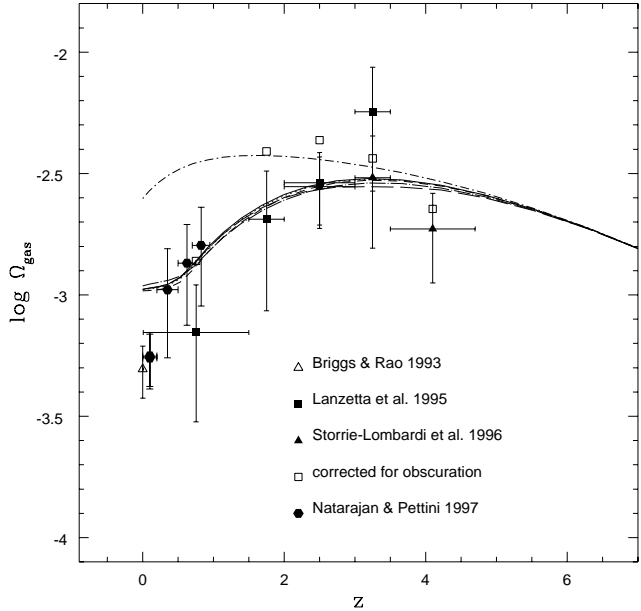,width=0.5\textwidth}
\caption{Evolution of the cold gas density parameter in damped Lyman--$\alpha$
absorbers. Solid squares: data without the APM QSO survey. Solid triangles:
data including the APM QSO survey. Open squares: tentative correction for
selection effects due to QSO obscuration (Storrie--Lombardi {\it et al.}
1996). Solid hexagons: Natarajan and Pettini (1997).
Open triangle: local estimate from HI surveys (Briggs and Rao
1993). Line codes of scenarios Q, A, B, C, D, E are given in tab. 1. The
various scenarios involving the ``burst'' mode consume more gas than the
``quiescent'' mode (dots and short dashes).}
\end{figure}

\begin{figure} 
\psfig{figure=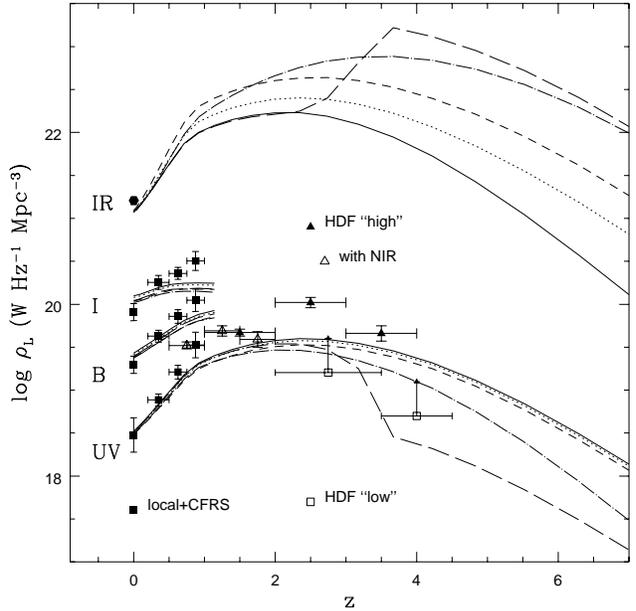,width=0.5\textwidth}
\caption{Rest--frame comoving luminosity density. Letters UV, B, I and IR
respectively stand for 2800\AA, 4400 \AA, 10000 \AA\ and 60 $\mu$m.  The
emissivity at 1600 \AA\ is about 30 \% higher than at 2800 \AA.  
The luminosity densities are based on the integration of the 
luminosity function fit on all magnitudes. Solid
squares: local and Canada--France Redshift Survey (Lilly {\it et al.}
1996). Open triangles: NIR data are taken into account to compute photometric
redshifts in the Hubble Deep Field (Connolly {\it et al.} 1997). Solid
triangles: other estimates of photometric redshifts in the HDF (Sawicki {\it
et al.} 1997). Open squares with arrows: 
HDF with redshifts from Lyman--continuum
drop--outs (Madau {\it et al.} 1996).  Solid hexagon: 60 $\mu$m local density
corresponding to one third of the bolometric light radiated in the IR
(Saunders {\it et al.} 1990).  Line codes of scenarios A, B, C, D, E are 
given in tab. 1. The different UV and IR emissions mainly result
from different fractions of ULIGs (with a top--heavy 
IMF and strong extinction), with almost similar SFR histories. Strongly
differring high--$z$ IR emission are obtained without being much constrained
by the current status of (discrepant) UV/optical observations.}
\end{figure}

\section{The history of star formation in the universe}

\subsection{The ``quiescent'' mode of star formation}

In this section, we will introduce scenarios of evolution which will be used
to compute the IR/submm properties of galaxies, and to predict faint galaxy
counts, and the CIB. The description of these scenarios and their line
codes in the figures are summarized in tab. 1. 
While a complete assessment of the energy budget of
galaxies would require the monitoring of multi--wavelength 
luminosity functions and galaxy
counts (delayed to forthcoming studies), we hereafter only wish to address the
issue of the overall evolution of the comoving SFR and gas densities in the
universe. Pei and Fall (1995) and Fall {\it et al.} (1996) emphasized the
relevance of these quantities to describe the overall evolution of the galaxy
population.

Fig. 9 and 10 respectively show the predicted SFR and gas evolutions for
scenario Q with $\beta=100$. It corresponds to the fit of SFR time scales in
disks, that is, the so--called ``Roberts times'' peaking at 3 Gyr and ranging
between 0.3 and 30 Gyr (Kennicutt {\it et al.} 1994).  The feedback parameters
are $\alpha=5$ and $V_{hot}=40(1+z_{coll})$ km s$^{-1}$.  This somewhat
arbitrary $(1+z)$ dependence is intended to mimic the effect of global
re-heating at high redshift and avoids a stepwise change of $V_{hot}$ at
$z=2$ caused by a bimodal regime. This has almost no effect at low $z$ but 
it changes the high--redshift behaviour of the SFR density.

Clearly the SFR density in scenario Q does not decline fast enough between
$z=1$ and the local universe, and, correspondingly, gas is not consumed
sufficiently. The impossibility to reproduce the peak of SFR density
with star formation histories similar to those in disks is not
surprising. We know that their
present SFR time scales (the Roberts times) are large and 
that their SFRs have not changed significantly during the last few 
billion years (Kennicutt {\it et al.} 1994). 

The prediction for the evolution of the comoving SFR density is very
similar to the result given by Baugh {\it et al.} (1997), though their model
involves a more accurate procedure to compute the merging history of haloes
(Cole {\it et al.} 1994). As in our model, they choose a SFR proportional to
the ratio of the cold gas content to the dynamical time.  But galaxies within
haloes can merge after spiralling to the centre of the haloes induced by
dynamical friction. During the merging, all the available gas is quickly
converted into stars. In spite of this explicit implementation of a kind of
``burst'' mode, starbursting does not seem to affect significantly the SFR
history, and the evolution they predict does not reproduce the steep decline
of the SFR density between $z=1$ and 0. Probably the modelling of dynamical
friction alone underestimates the true magnitude of the interaction and merging
processes between galaxies in merging haloes.

\begin{figure} 
\psfig{figure=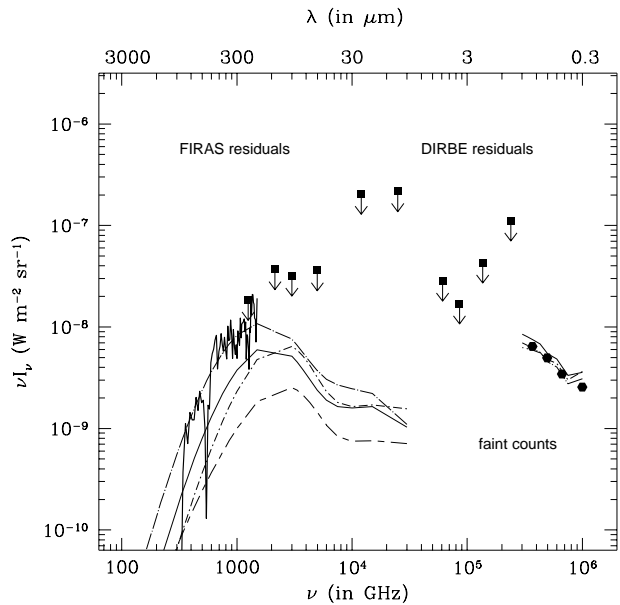,width=0.5\textwidth}
\caption{Predictions of the diffuse backgrounds in the FIR/submm and in the
optical compared to the current upper limits and detections.  The solid
squares show the level of {\em COBE}/DIRBE residuals from Hauser (1995).  The
similar shapes of the residuals and dark sky suggest that the subtraction of
foregrounds has been incomplete and that the plotted values are upper limits.
The thick solid line gives the CIB spectrum from
the re--analysis of the {\em COBE}/FIRAS residuals, initiated in 
Puget {\it et al.}
(1996) and revisited in Guiderdoni {\it et al.} (1997).  The solid hexagons
show the Cosmic Optical Background (COB) obtained by summing up faint galaxy
counts down to the Hubble Deep Field limit. Strictly speaking, this is only a
lower limit of the actual COB, but the shallowing of the $U$ and $B$--band
counts suggests near--convergence at least at those wavelengths (Williams {\it
et al.} 1996).  The short dashes and long dashes give the prediction for
no--evolution integrated up to redshift $z_{for}=8$ in a cosmology with
$h=0.5$ and $\Omega_0=1$.  The other curves are computed for the SCDM model
with $h=0.5$, $\Omega_0=1$, $\sigma_8=0.67$, for scenarios Q, A, and E plotted
with the line codes of tab. 1. Scenarios Q and A are not sufficient to 
reproduce the CIB.  This suggests the existence of an
additional population of ULIGs taken into account in scenario E.}
\end{figure}

\subsection{The ``burst'' mode of star formation}

We now wish to introduce another mode of star formation with $\beta=10$.  The
SFR time scales are now much shorter. As a result, the evolution of the 
luminosity function
differs from the evolution in scenario Q involving the ``quiescent'' mode of star
formation. As shown in fig. 2, the luminosity function of the ``quiescent'' 
mode is built up by the
accumulation of light coming from all galaxies, and its evolution reproduces 
the evolution of the mass distribution of collapsed haloes. On the contrary, 
the luminosity function of the ``burst'' mode reproduces the evolution 
in the formation of
haloes. Each generation of halo formation is accompanied by a strong starburst
which acts as a transient beacon. As a consequence, the evolutionary trends of
the luminosity functions in the two modes strongly differ.

Since we know that the local universe is dominated by the ``quiescent'' mode, we
now consider a mix of two broad types of populations, one with a ``quiescent''
star formation rate, the other proceeding in bursts.  For the ``burst'' mode,
we take an involved mass fraction increasing with redshift
$f_{burst}(z)=f_{burst}(0)(1+z_{coll})^\gamma$, as suggested by the increasing
fraction of blue objects showing tidal and merger features at larger $z$
(Abraham {\it et al.} 1996). Such a dependence is also consistent with
theoretical considerations on the merger rates of galaxy pairs in
merged haloes (Carlberg 1990). We limit the scope of the present paper 
to this phenomenological modelling of $f_{burst}(z)$.
This is clearly a point which should be refined in forthcoming studies.
Noting that the frequency of galaxy pairs is
$\propto (1+z)^\delta$ with $\delta$ between 2 and 6 at $\pm 1 \sigma$ (Zepf
and Koo 1989; Burkey {\it et al.} 1994; Carlberg {\it et al.} 1994), we choose
here a high evolution rate $\gamma =5$. Then $f_{burst}(0)$ is set to 0.04 in
order to fit the SFR density at low $z$, resulting in an ``all--burst''
behaviour at $z_{coll} \geq 0.8$.  This will be our scenario A.  As it appears
from fig. 9, this phenomenological description of the increasing importance of
bursts reproduces the steep decline of the SFR density. The origin of this
fast evolution has still to be elucidated by a more exhaustive modelling of
all interaction processes in the semi--analytic codes which follow the merging
history trees of haloes and galaxies.  Fig. 9 also shows that the predicted
cosmic SFR history has a high--$z$ evolution closer to the one derived from
photometric estimates of redshifts rather than that from Lyman--continuum
drop--outs. For instance, one should note that at $z=4$, the predicted SFR is
ten times the lower limit derived from the Lyman--continuum drop--outs
{\it without extinction}. As a consequence of the
high SFR, the gas content strongly evolves between $z \sim 2$ and 0, as shown
in fig. 10.

\begin{figure} 
\psfig{figure=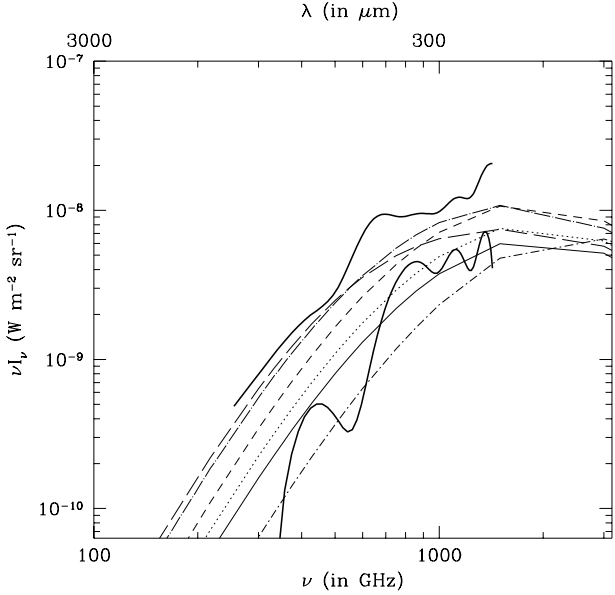,width=0.5\textwidth}
\caption{Predictions of the diffuse backgrounds in the FIR/submm (blow up),
compared to the acceptable range of the CIB at $\pm 1\sigma$ per point
(Guiderdoni {\it et al.} 1997).
Scenarios Q, A, B, C, D, and E are plotted with line codes of tab. 1.
The predictions for the COB are similar to those in fig. 12.}
\end{figure}

We can now compute the corresponding IR/submm emission by taking the
conservative estimate of the average optical thickness and ``slab'' geometry
as in the ``quiescent'' mode. As a result, this population of ``mild starbursts''
and ``luminous UV/IR galaxies'' (LIGs) have IR--to--blue luminosity ratios in
the range $0.06 \leq L_{IR}/\lambda_B L_B \leq 4$ which is characteristic of
blue--band selected samples (Soifer {\it et al.}  1987), and should be fitted
to the Canada--France Redshift Survey (selected in the observer--frame
$I_{AB}$ band, roughly corresponding to the $B$ band at $z \sim 1$), and to
high--$z$ HDF galaxies. Galaxies at the peak of the SFR time--scale
distribution ($t_\star=0.3$ Gyr) have $L_{IR}/M \simeq 6.3
L_{bol\odot}/M_\odot$.  Nevertheless, their colours can still be very blue
during the burst ($B-V=0.1$ at 0.5 Gyr).  The evolution of the comoving
luminosity density in various UV/visible bands and at 60 $\mu$m is compared to
observational estimates in fig. 11. The local energy budget and its evolution
from $z=0$ to 1 seem to be fairly reproduced. The difference between the
fits of the comoving SFR and luminosity densities (fig. 9 and 11) originates
in the presence of dust absorption. The local UV flux is correctly fitted in
fig. 11, while the local SFR in fig. 9 is about twice the value deduced from
the local UV flux {\it under the assumption of no extinction}.

By integrating the contribution of sources along the line of sight, we can
predict the corresponding diffuse backgrounds.  While the results for both
scenarios fairly reproduce the COB, as shown in fig. 12, scenario A does not
do much better than scenario Q to reproduce the CIB.  Their predictions are
clearly barely compatible with the acceptable $\pm 1 \sigma$ range recalled in
fig. 13.  The mean amplitude of the CIB is twice the prediction,
despite our high choice of $\gamma$. Consequently, we can strongly suspect the
existence of a population of galaxies which are more heavily extinguished.

\begin{figure}
\psfig{figure=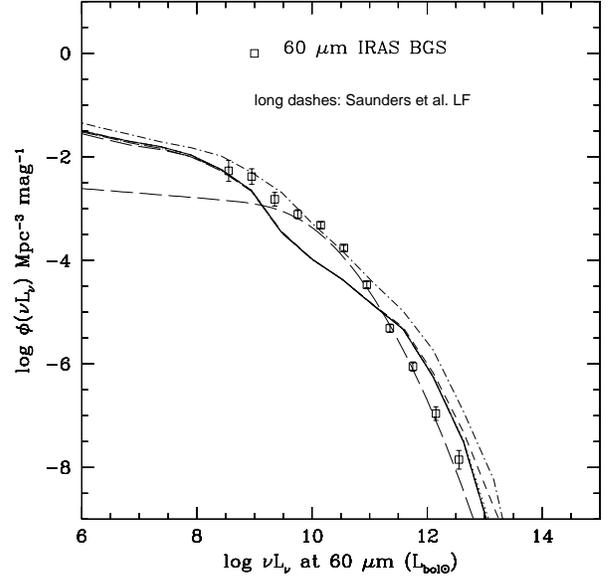,width=0.5\textwidth}
\caption{Predicted 60 $\mu$m luminosity functions at $z=0$ for the SCDM model
with $h=0.5$, $\Omega_0=1$ and $\sigma_8=0.67$. 
Scenarios Q, A, B, C, D, and E are plotted with line codes of tab. 1. 
At $z=0$, scenarios A, B, C, D, E give
almost similar results since the ``burst'' mode only involves 4 \% of the
mass.  Scenario Q gives a slightly higher luminosity function since more 
gas is left
to fuel star formation at $z=0$.  Squares: observational luminosity 
function for the {\em IRAS}
Bright Galaxy Sample (Soifer and Neugebauer 1991). Thick long dashes:
observational luminosity function from a compilation of various samples
(Saunders {\it et al.}  1990).}
\end{figure}

\subsection{A heavily--extinguished component}

The above-mentionned CIB computed with scenario A seems to be something like a
conservative estimate of the minimum IR/submm background due to typical CFRS
and HDF galaxies.  We now wish to assess how much star formation might be
hidden by dust shrouds and introduce an additional population of
heavily--extinguished bursts, which are similar to ``ultra-luminous IR
galaxies'', or ULIGs (Sanders and Mirabel 1996; Clements {\it et al.}
1996a). We maximize their IR luminosity by assuming that all the energy
available from stellar nucleosynthesis ($0.007xMc^2$) is radiated in a
heavily--extinguished medium, yielding a total IR luminosity $L_{IR}/M \simeq
42.5(x/0.40)(t_\star/1Gyr)^{-1}$ $L_{bol\odot}/M_\odot$, provided the
lifetimes of stars are smaller than the duration $t_\star$ of the burst. We
take $<x>=0.40$ for stars with masses larger than $\sim 5 M_\odot$ (Schaller
{\it et al.} 1992), and get luminosities which are a factor $\sim 20$ larger
than those of our mild--starburst/LIG mode.  As a consequence of this simple
model, the starburst only radiates in the IR/submm, and only dark remnants
are left when it stops, without the slow evolution of low--mass stars.
We distribute this population of
ULIGs in two ways: i) A constant mass fraction of 5 \% (scenario B) or 15 \%
(scenario D) at all $z_{coll}$, mimicking a scenario of continuous bulge 
formation as
the end--product of interaction and merging.  ii) 90 \% of all galaxies
forming at high $z_{coll} \geq 3.5$ undergo a heavily--extinguished burst,
mimicking a strong episode of bulge formation. These scenarios are now able to
fit both the COB and CIB. Scenarios B and D seem more appropriate to
reproducing the CIB at 300 $\mu$m while scenario C, with high--$z$ ULIGs, has
a stronger contribution at larger wavelengths. Of course, these last three
cases are only illustrative, and a combination of these solutions would also
fit the CIB. For instance, we introduce an ad hoc scenario E with a fraction
of ULIGs increasing as $1- \exp -0.02(1+z_{coll})^2$. Such a dependence can be
obtained if the fraction of ULIGs depends on the mean surface density and
optical thickness of disks which roughly scale as $(1+z_{coll})^2$ in our
modelling.  As shown by fig. 12 and 13, scenario E nicely reproduces 
the COB and CIB, and could be considered as our ``best fit''.

It is clear from fig. 11 that none of the optical data reflects the large
differences between these scenarios, although the fraction of light in the IR
varies widely at high $z$.  In scenario A, the IR/UV ratio decreases with
increasing $z$ because of the decreasing metallicity of galaxies. In scenario
B and D, this effect in cancelled because the ULIG bursts are assumed to be
optically--thick, with a top--heavy IMF, and the IR/UV ratio at $z=4$ is
similar to that at $z=0$.  In scenario C and E, the IR/UV ratio strongly
increases with $z$ and is $\sim 100$ times higher than for model B at
$z=4$. One should note that at this redshift, scenario C and E are roughly
consistent with the lower limits of the UV--luminosity density derived from
Lyman--continuum drop--outs, but with ten times as much SFR as directly
derived if extinction is not taken into account. In these scenarios, galaxy
formation at high $z$ is an almost completely--obscured process.

We emphasize that the family of scenarios summarized in tab. 1 is introduced
within the same SCDM model. The dissipative and non--dissipative collapses are
unchanged, and the characteristic time scale for the conversion of gas into
stars is always proportional to the local dynamical time, which is the most
natural time scale.  Given that, we only change the ``fuzzy'' astrophysics
introduced with the efficiency parameter $\beta$. It seems reasonable to {\it
assume} that the efficiency of star formation is typically an order of
magnitude greater for interacting galaxies (the ``burst'' mode) than for
isolated galaxies (the ``quiescent'' mode), resulting in a $\beta$ parameter an
order of magnitude lower. While this assumption has yet to be put on firmer
grounds, its effect on the global SFR and gas densities is very strong.  As a
second step, changes in the IMF and extinction can modify the optical and
IR/submm energy budget.

\begin{figure}
\psfig{figure=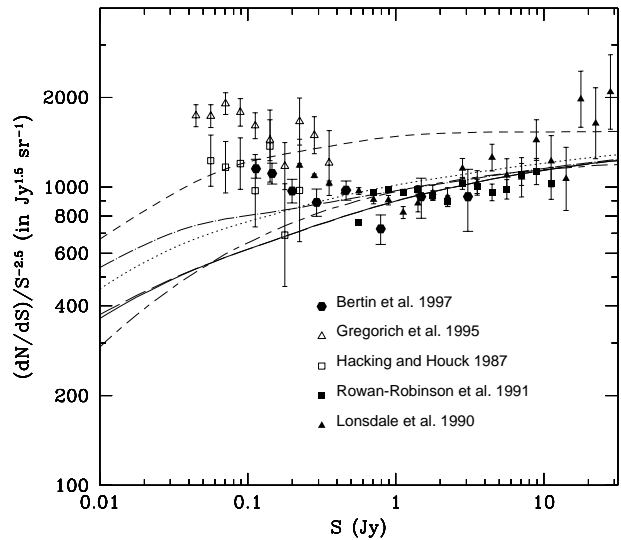,width=0.5\textwidth}
\caption{Predictions for differential galaxy counts (normalised to Euclidean
counts) at 60 $\mu$m. Data is shown for {\em IRAS} counts at 60 $\mu$m. Solid 
triangles:
Faint Source Survey (Lonsdale {\it et al.} 1990). Solid squares: QMW survey
(Rowan--Robinson {\it et al.} 1991b). Solid hexagons: revisited counts in the
Very Faint Source Survey (Bertin {\it et al.} 1997).  Open squares: North
Ecliptic Pole Region (Hacking and Houck 1987). These latter counts might be
affected by the presence of a super--cluster.  The counts by Gregorich {\it et
al.}  (1995), which are plotted with open triangles, are probably
contaminated by
cirrus.  The no--evolution model (for a cosmology with $h=0.5$, $\Omega_0=1$)
is shown with short dashes and long dashes. Scenarios A, B, C, D, E are plotted
with line codes of tab. 1. Scenario D has too many local ULIGs and should
be rejected. Scenario E with an increasing fraction of ULIGs almost reproduces
the flat behaviour of the counts.}
\end{figure}

\section{Predictions in the FIR and submm}

\subsection{The FIR luminosity function}

Fig. 14 gives predictions for the $z=0$ luminosity function at 60 $\mu$m
compared with the observational determinations drawn from the {\em IRAS} Bright
Galaxy Sample (Soifer and Neugebauer 1991) and from a compilation of various
{\em IRAS} samples (Saunders {\it et al.} 1990).  The difference of the two
observational luminosity functions at the faint end illustrates the 
uncertainties. The agreement
of the predictions with the data is fair. It is not surprising that scenario Q
seems to give a better fit, since most galaxies of the BGS are spirals, mild
starbursts and LIGs.  All the scenarios involving the burst mode (A to E) give
similar results at $z=0$ since the fraction of mass involved in starbursts is
low (4 \%). Burst scenarios with shorter SFR time scales in the past consumed
more gas than scenario Q and less fuel is now left for star formation.

It has been recalled in Sec. 2.6 that this kind of semi--analytic models
predicts too many low--luminosity galaxies in the optical bands at $z=0$, with
respect to the observational field luminosity functions (see the references
quoted in Sec. 2.6).  As a consequence, the optical luminosity functions are
too steep and can be reconciled with the observations by invoking subtle
selection effects based on surface brightness.  
Our model also predicts too many low--luminosity galaxies
in the IR, while the discrepancy is somewhat lower than in the blue. The slope
is $\phi(L_{IR})dL_{IR} \propto L_{IR}^{-1.25}dL_{IR}$.  The normalisation at
the bright end can be improved by a slight decrease (a few tenth dex) of the
baryonic density $\Omega_{bar}$, or a slight increase of the baryonic
mass--to--luminosity ratios, through the normalisation of the ``dark mass'' in
the IMF. We do not attempt to apply this kind of fine tuning.

\begin{figure}
\psfig{figure=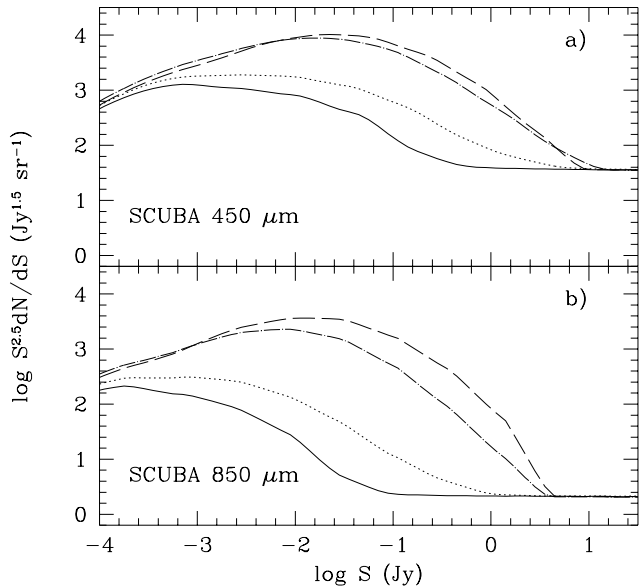,width=0.5\textwidth}
\caption{Predictions for differential galaxy counts (normalised to Euclidean
counts) at 450 and 850 $\mu$m for observations with SCUBA. 
In contrast with counts at wavelengths smaller
than 100 $\mu$m, the submm counts are very sensitive to the details of the
high--$z$ evolution, because of the shift of the 100 $\mu$m bump into the
observing bands.  Scenarios A, B, C and E are plotted with line codes of 
tab. 1.}
\end{figure}

\subsection{Faint galaxy counts}

We hereafter explore the faint counts predicted by using the scenarios
proposed in tab. 1.  Fig. 15 gives the predictions for the {\em IRAS} 60 $\mu$m
differential counts normalised to the Euclidean slope. Observational counts
from the QMW survey (Rowan--Robinson {\it et al.} 1991b), the Faint Source
Survey (Lonsdale {\it et al.} 1990), the Very Faint Source Survey (Bertin {\it
et al.} 1997), and the North Ecliptic Pole Region (Hacking and Houck 1987) are
superimposed to the predictions. The latter survey might be partly affected by
a large--scale structure at $z=0.088$ (Ashby {\it et al.} 1996). As shown by
Bertin {\it et al.} (1997), the counts by Gregorich {\it et al.} (1995) are
contaminated by cirrus. The theoretical curves are {\it not} renormalised to
the bright end of the observed counts. The scenarios involving an increasing
fraction of the ``burst'' mode predict more galaxies than the no--evolution
curve.  Scenario D with 15 \% of ULIGs is rejected by the data. All the other
scenarios have a moderate local fraction of ULIGs and are in agreement with
the faint counts. Scenario E with an increasing fraction of ULIGs gives an
almost flat curve which is reminiscent of the observational trend.
Nevertheless, the rise of the counts below 0.1 Jy, if it is real, is not
reproduced. However, our scenarios predict a correct amount of fluctuation and
there is not much space for stronger evolution. More specifically, the counts
predict a 60 $\mu$m background fluctuation per beam in the Very Faint Source
Survey (after removal of $\geq 4\sigma_{tot}=120$ mJy sources) at the level of
14.1 mJy (A), 16.0 mJy (B), 14.3 mJy (C) and 17.0 mJy (E), while the measured
68 \% quantile is $30.1 \pm$ 1.2 mJy (Bertin {\it et al.} 1997).  With a 25
mJy {\it rms} instrumental noise and 6.5 mJy {\it rms} cirrus fluctuations
(Gautier {\it et al.} 1992), there is still space for a $15.4^{+2.2}_{-2.5}$
mJy fluctuation due to sources, in good agreement with our estimates.
Moreover, if the cirrus fluctuations are non--gaussian, the {\it rms} value
strongly overestimates the 68 \% quantile and can tolerate
$16.8^{+2.0}_{-2.3}$ mJy for source fluctuation.

Fig. 16 and 17 give the predictions for various wavelengths from 15 $\mu$m to
1.4 mm which correspond to current and forthcoming instruments.  The {\em IRAS} 60
$\mu$m counts are recalled in one of the panels, as well as the ISO--HDF 15
$\mu$m deep counts (Oliver {\it et al.} 1997).  Only scenarios A and E are
plotted. Clearly there are three regimes: (i) Nearby galaxies are in the
Euclidean zone, giving a count slope $N(>S) \propto S^{-3/2}$. The value of
the bright--end normalisation at submm wavelengths depends on the choice of
the emissivity index $m$.  Since the count rates are very low, it is likely
that we shall have to wait for the all--sky survey of {\em PLANCK} in order to fix
the counts at the bright end.  (ii) At short wavelengths, the curvature effect
and the positive ``k--correction'' produce the bend of the faint counts; (iii)
In the submm/mm range, the negative ``k--correction'' produces a bump in the
faint counts, which reflects the passage of the 100 $\mu$m emission bump into
the observing bands.  

It appears from fig. 17 that the ISO--HDF counts suggest evolution,
but that the predictions are not strongly sensitive to the details of
the evolutionary scenario.
Guiderdoni {\it et al.} (1997) have emphasized the relative degeneracy of the
predictions at wavelengths shorter than 100 $\mu$m, including the {\em IRAS} 60
$\mu$m and the {\em ISO} 15 $\mu$m counts, and the strong sensitivity of the submm
counts to the details of galaxy evolution, which is already apparent for 
{\em ISO} observations at 175 $\mu$m.  Tab. 2 and 3 gather the count
predictions for the various scenarios of tab. 1, and two typical sensitivity
levels: 100 mJy which will be reached for point sources by the {\em PLANCK} all--sky
survey, and which is comparable to the 0.22 Jy level of the {\em IRAS} 60 $\mu$m
Faint Source Catalogue; and 10 mJy which can be reached by a deep pencil--beam
survey with SCUBA and {\em FIRST}. The number counts at 100 mJy and 10 mJy are very
sensitive to evolution and can differ by an order of magnitude at 350 $\mu$m,
and more than two orders of magnitude at 850 $\mu$m.  This explains the
discrepancy of the predictions in the literature so far published. Hivon {\it
et al.} (1997) will come back to the detection of submm sources and of the
fluctuations they induce.  As a conclusion of these predictions, while the
{\em IRAS} data do not yield tight constraints on the evolution of galaxies at $z
\geq 0.2$, the forthcoming deep survey with {\em ISO} at 175 $\mu$m and with SCUBA in
the atmospheric windows at 450 and 850 $\mu$m (for which the array is matched
to the instrument optics) will quickly help discriminate between the various
scenarios, in the expectation of {\em PLANCK} and {\em FIRST}.

\begin{figure*}
\psfig{figure=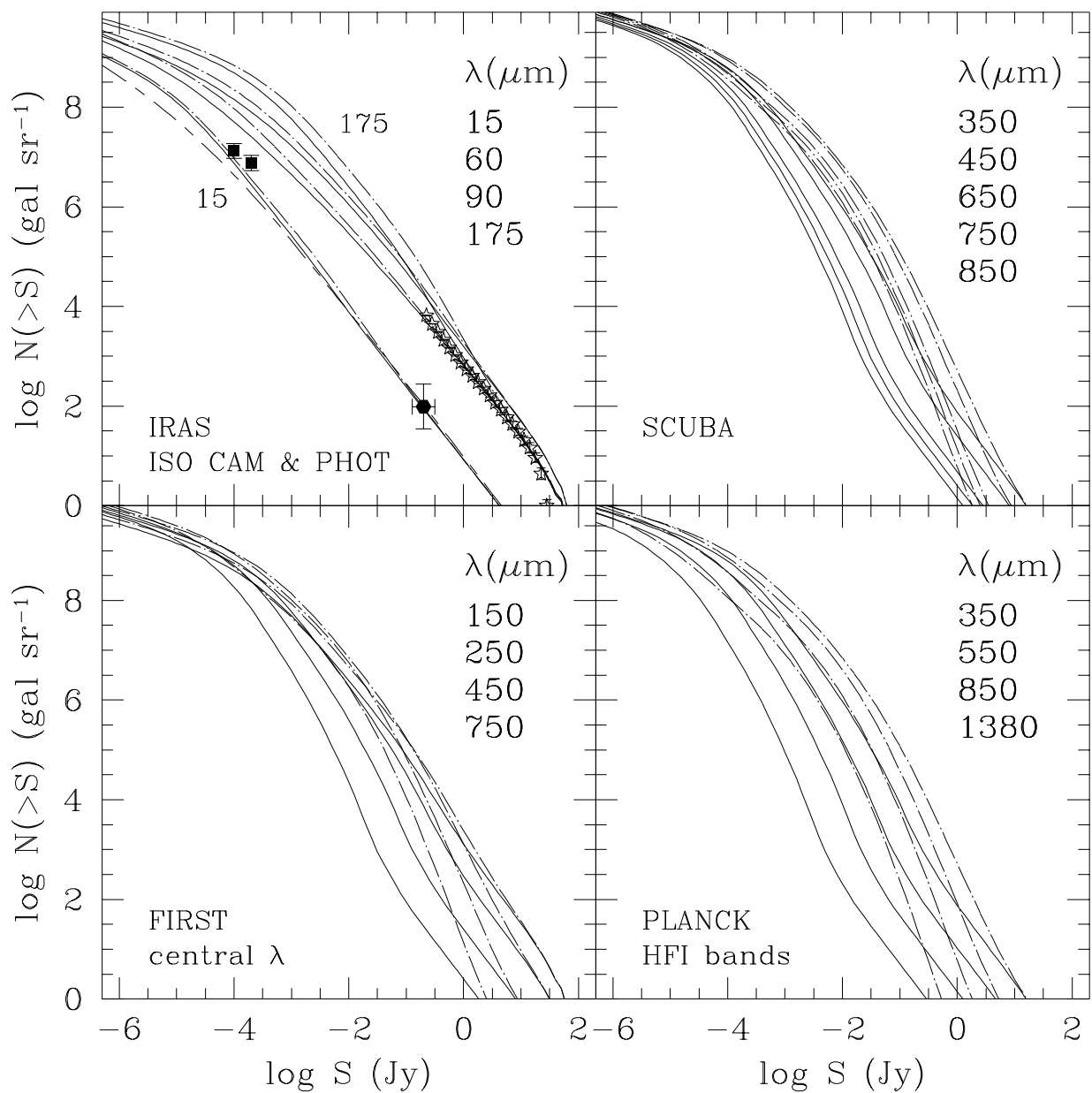,width=\textwidth}
\caption{Predictions for number counts at various wavelengths for operating
and forthcoming observing facilities and experiments: ISOCAM and ISOPHOT
on--board {\em ISO}, SCUBA, {\em FIRST}, and {\em PLANCK} High Frequency Instrument.  Some of
the curves are redundant. The curves correspond to the wavelengths from top to
bottom, except in the {\em IRAS}/{\em ISO} Panel where it is from bottom to top. Note that
the 15 $\mu$m curve only takes into account dust emission (without the stellar
component), and the observer--frame fluxes should be considered as lower
values beyond $z \sim 1$. The no--evolution prediction at 15 $\mu$m
(for $h=0.5$ and $\Omega_0=1$) is plotted with short dashes and long dashes. 
Open stars: {\em IRAS} at 60 $\mu$m (Lonsdale {\it et
al.} 1990).  Solid hexagon: {\em IRAS} at 12 $\mu$m (Rush {\it et al.} 1993).  Solid
squares: ISO--HDF counts at 15 $\mu$m (Oliver {\it et al.} 1997). Scenarios A
and E are plotted with line codes as in tab. 1.}
\end{figure*}

\subsection{Redshift distributions}

Fig. 18 shows the prediction for the redshift distribution corresponding to
the spectroscopic follow--up of two {\em IRAS} samples: the North Ecliptic Pole
Region observed by Ashby {\it et al.} (1996) with 60 $\mu$m fluxes $60 \leq
S_{60} \leq 150$ mJy, and the sample of Clements {\it et al.} (1996a) for
galaxies of the Faint Source Catalogue with $S_{60} \geq 300$ mJy which are
`IR loud'' (IR--to--blue ratios larger than 10). The data of the NEPR are
almost complete and can be directly compared to our predictions. Nevertheless,
the first two bins are probably affected by a super--cluster at $z=0.088$. The
redshifts of the ``IR--loud'' FSC sample have been taken by the authors in
order to extract a subsample of ULIGs and not for a complete redshift survey.
The histogram given for the galaxies which are not ULIGs (144 sources) has
been roughly rescaled to take into account the number of sources not quoted in
their Appendix because they already have redshifts in the literature.  The
shaded histogram shows their 91 ULIGs. Then the histogram is renormalised in
order to peak at the same relative level as the NEPR. The sample has only a 60
\% completeness. In general, the scenarios seem to predict more galaxies at
high $z$ than in the NEPR sample. The ``IR--loud'' FSC sample plotted here
suffers from problems of overall normalisation and incompleteness, but it
signals the presence of galaxies at $z \geq 0.2$. The difference between
the redshift distributions for scenarios E and A gives the fraction of 
ULIGs in scenario E (there are no ULIGs in scenario A). The distribution 
of ULIGs with $z$ seems to be much flatter than the
observed peak in the ``IR--loud'' FSC sample. However, it is
difficult to assess the level of discrepancy with a 60 \% the incompleteness. 
Clearly the redshift follow--up of deep IR--selected samples is still an 
on--going project, and it will bring crucial information on the nature
of the sources.

 Fig. 19 shows predictions for the redshift distributions at 60, 200, 350, and
550 $\mu$m, corresponding to various flux depths.  Scenarios A and E (as well
as B, C, and D not plotted here) give different predictions at $z \geq 1$. For
scenario A, the redshift distributions at the 10 mJy flux level peak at $z
\sim 1$, with a high--$z$ tail encompassing 90 \% of the sources at $z \sim
2.5$ to 3. Scenario E has galaxies at still higher $z$. In contrast with the
{\em IRAS} samples at 60 $\mu$m which, at the 100 mJy flux level, only probe the
universe at $z \sim 0.2$, the future spectroscopic follow--up of
submm observations at the 10 mJy level should
be able to explore the IR/submm evolution of galaxies at $z \geq 1$.  Finally,
fig. 20 shows the contribution of sources fainter than $S$ to the background
value at 60, 200 and 550 $\mu$m. The 10 mJy level which will be reachable by
the forthcoming submm observations will allow the surveys to begin
``breaking'' the CIB into discrete units.

\begin{table}
\caption{Predictions of galaxy counts in {\em PLANCK} HFI bands at the 100 mJy level
(log number of sources sr$^{-1}$).}
\begin{center}
\begin{tabular}{@{}lrrrr} \hline
Name & 350 $\mu$m & 550 $\mu$m & 850 $\mu$m & 1380 $\mu$m \\ Q & 3.97 & 3.06 &
2.15 & 1.18 \\ A & 3.93 & 2.73 & 1.67 & 0.71 \\ B & 4.41 & 3.48 & 2.23 & 0.83
\\ C & 4.95 & 4.88 & 4.39 & 3.15 \\ D & 5.04 & 4.34 & 3.33 & 1.77 \\ E & 5.08
& 4.57 & 3.83 & 2.39 \\
\end{tabular}
\label{ta:planck}
\end{center}
\end{table}

\begin{table}
\caption{Predictions of galaxy counts in SCUBA and {\em FIRST} bands at the 10 mJy
level (log number of sources sr$^{-1}$).}
\begin{center}
\begin{tabular}{@{}lrrrrr} \hline
Name & 350 $\mu$m & 450 $\mu$m & 650 $\mu$m & 750 $\mu$m & 850 $\mu$m \\ Q &
5.66 & 5.23 & 4.41 & 4.09 & 3.80 \\ A & 5.98 & 5.63 & 4.73 & 4.40 & 3.98 \\ B
& 6.27 & 5.99 & 5.34 & 5.07 & 4.75 \\ C & 6.80 & 6.81 & 6.64 & 6.52 & 6.37 \\
D & 6.66 & 6.47 & 6.01 & 5.80 & 5.55 \\ E & 6.84 & 6.75 & 6.45 & 6.29 & 6.09
\\
\end{tabular}
\label{ta:scuba}
\end{center}
\end{table}

\begin{figure}
\psfig{figure=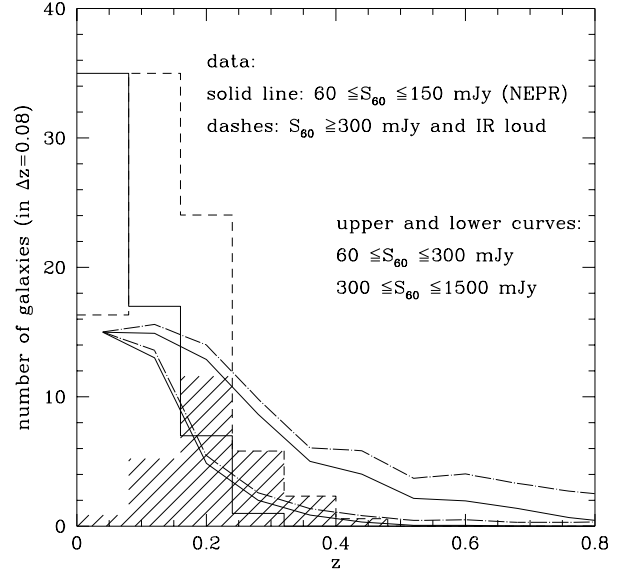,width=0.5\textwidth}
\caption{Predicted redshift distribution of the North Ecliptic Pole Region
observed by Ashby {\it et al.} 1996. The observed galaxies have 60 $\mu$m
fluxes in the range $60 \leq S_{60} \leq 150$ mJy.  The first two bins might
be affected by a super--cluster.  The sample of Clements {\it et al.} 
(1996a) is
also plotted for sake of illustration.  The reader is referred to this paper
for the description of the selection criteria. Roughly, the galaxies have
$S_{60} \geq 300$ mJy and are ``IR loud'' (IR--to--blue ratio larger than
10). See text for the details. The shaded histogram show the ULIGs of this
sample. Scenarios A and E are plotted with line codes of tab. 1. Their
normalisation is relative.}
\end{figure}

\section{Discussion and conclusions}

We have used a simple semi--analytic model of galaxy formation to derive the
IR/submm statistical properties of galaxies: luminosity function, faint galaxy
counts, redshift distributions, and the diffuse background. The model shares
many common features with previous semi--analytic works focussing upon the
optical properties of galaxies: (i) the collapse of the perturbations is
described by the classical top--hat model under the assumptions of homogeneity
and sphericity; in this simple version, we used the peaks formalism to compute
the formation rate of haloes. (ii) the dissipative cooling and collapse is
introduced, with the usual ``overcooling'' problem which can be partly
suppressed by introducing feedback due to overall re--ionization and
galactic winds triggered by SNe.  (iii) star formation is proportional to the
ratio of the gas content to the dynamical time scale of the galaxies; and (iv)
stellar evolution is explicitly implemented.

In order to make specific predictions for the IR/submm wavelength range, this
model includes the following assumptions:

\begin{enumerate}

\item An average optical depth of the disks is implemented and scales as the
relative gas content and gas metallicity (estimated through the assumption of
Instant Recycling Approximation); no evolution of the dust composition is
considered; the variation of the extinction curve with metallicity
is based on an interpolation of the Milky Way and the Magellanic
Clouds which gives an extinction proportional to $Z^{1.6}$;

\item A simple ``slab'' geometry is assumed where dust and stars are mixed
with equal height scales;

\item Finally, the dust emission spectrum scales with the bolometric IR
luminosity, following the observational relationship of {\em IRAS} colours with
luminosities. Such a dependence is obtained by changing the abundance of three
components: PAH, very small grains and big grains (and the temperature of the
big grains).

\end{enumerate}

\begin{figure*}
\psfig{figure=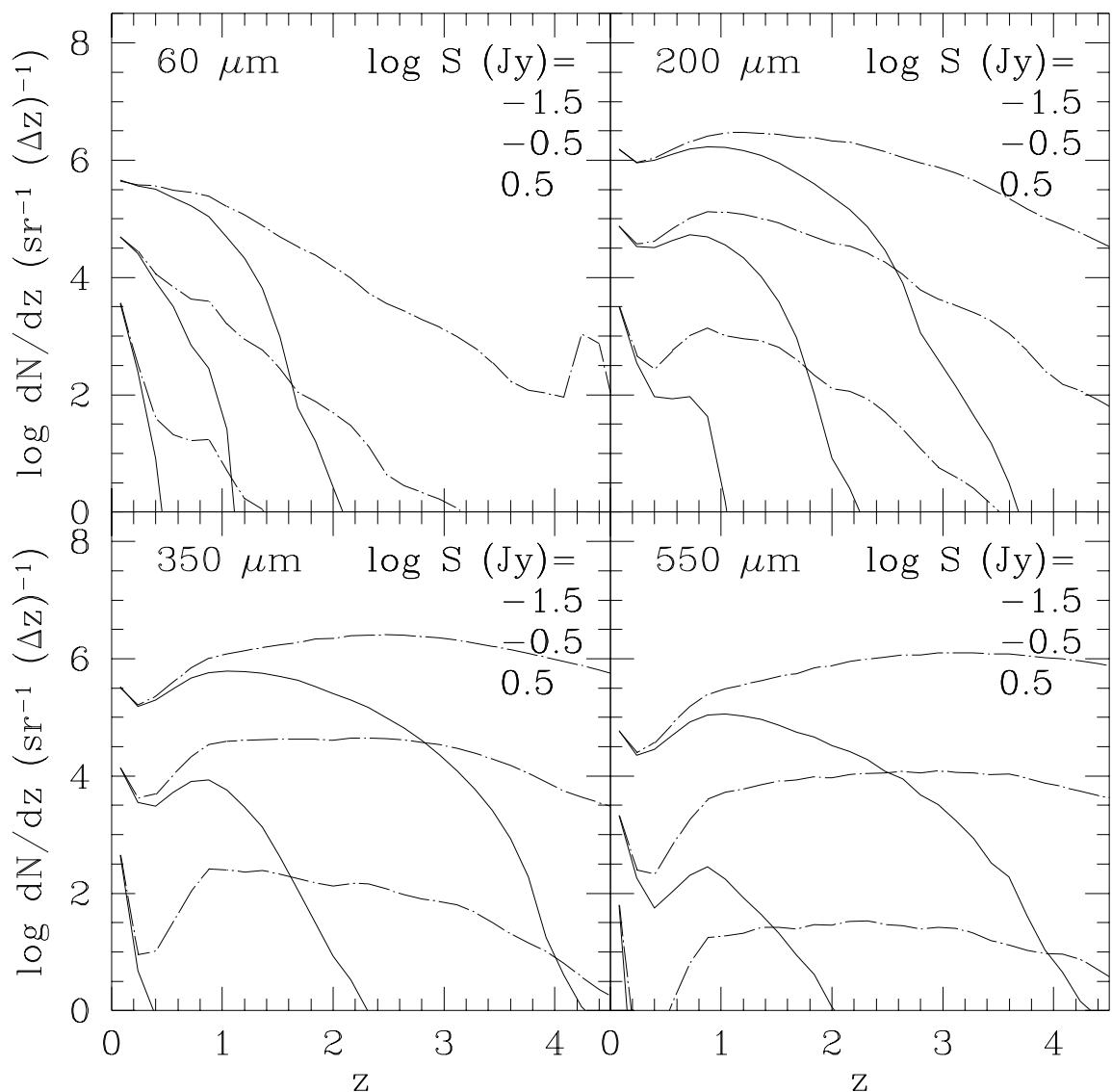,width=\textwidth}
\vskip -1truecm
\caption{Predictions for redshift distributions at four wavelengths, and in
different flux ranges. The flux bin widths are $\Delta \log S = 1$, and
the central flux values ($\log S$ in Jy) are shown in each figure 
(from top to bottom).
Scenarios A and E are plotted with line codes of tab. 1.}
\end{figure*}

We introduced two modes of star formation. In the ``quiescent'' mode'', the
distribution of the characteristic time scales $t_\star$ for star formation
peaks at 3 Gyr, and ranges from 0.3 to 30 Gyr. In the ``burst mode'', the
distribution of $t_\star$ peaks at 0.30 Gyr, and ranges from 0.03 to 3 Gyr.
The ``quiescent'' mode is unable to reproduce the rapid evolution of the cosmic
SFR and gas densities. The steep decline of the SFR and gas content since
$z=2$ can be accommodated by assuming an increasing fraction of mass involved
in the ``burst'' mode. Since one knows that the fraction of peculiar objects
with signs of interaction/merging increases with the depth of the survey, we
suspect that the bursting behaviour of the CFRS and HDF galaxies is due to
interactions and we introduce a phenomenological $z$ dependence fitting the
evolution of the pair rates.  Since LIGs and ULIGs are
respectively interacting and merging systems, it is reasonable
to estimate that, reciprocally, CFRS and HDF galaxies with
peculiar morphologies should emit in the IR. For these ``mild starbursts'' and
``luminous IR/UV galaxies'', we assume a conservative range of IR--to--blue
luminosity ratios $0.06 \leq L_{IR}/\lambda_B L_B \leq 4$ computed from our
standard assumptions on average optical thickness and geometry. Finally, in
order to reproduce the bright end of IR galaxies with $L_{IR}/\lambda_B L_B
\geq 10$, we also introduced a population of heavily--extinguished galaxies
with massive star formation, similar to ULIGs.

With these ingredients, we designed a family of evolutionary scenarios which
fit the local overall energy budget and its evolution to $z \sim 1$.  These
scenarios also fit the COB estimate computed from faint galaxy counts, and are
consistent with the observational range for the CIB. So we feel confident that
we roughly reproduce the local energy budget as well as a plausible range for
the integrated budget along the line of sight.  The scenarios predict UV and
IR emissions strongly differring at high $z$. With the same SFR history,
different UV fluxes can be obtained with different extinctions and IMF.  In
some of the scenarios, the UV fluxes at $z=4$ fit the current measurements (or
lower values) of Madau {\it et al.} (1996), but with ten times more SFR than
deduced {\it under the assumption of no extinction}. Such a factor 10,
which includes the contribution of heavily--extinguished ULIGs, is much 
larger than the factor 3 corresponding to the correction for extinction 
derived by Pettini {\it et al.} (1997) for the HDF galaxies with 
Lyman--continuum drop--outs. This quantifies the
strong warning that the deduction of the SFR density from UV fluxes misses
the optically--dark side of galaxy formation. 

\begin{figure}
\psfig{figure=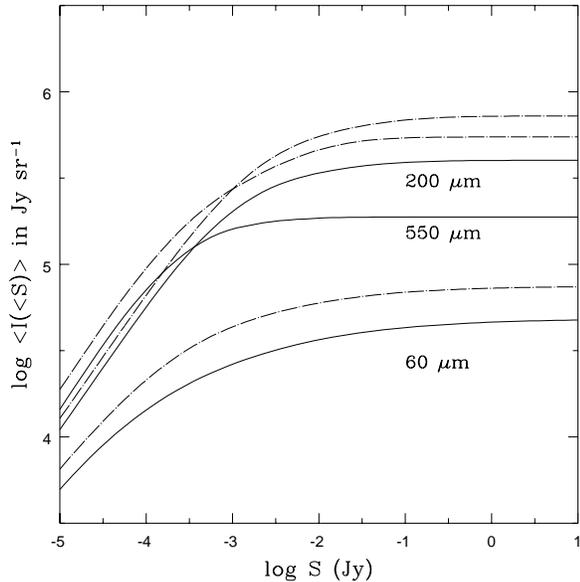,width=0.5\textwidth}
\caption{ Contribution of sources fainter than $S$ to the background value at
60, 200 and 550 $\mu$m.  Scenarios A and E are plotted with line codes of
tab. 1.}
\end{figure}

This family of scenarios was obtained within the context of the same
cosmological model, with similar prescriptions for dissipative and
non--dissipative collapses, star formation and feedback. The changes only
affect the efficiency of star formation in a dynamical time scale (the
$\beta$ parameter), the IMF (Salpeter or only massive stars), and the
extinction (average extinction with ``slab'' geometry, or strong,
``screen''--like extinction). We saw the extreme sensitivity of the
predictions to the parameter choice for this poorly constrained astrophysics.

The scenarios are consistent with {\em IRAS} data. They give similar predictions for
the {\em IRAS} 60 $\mu$m luminosity function and faint counts, provided the local
fraction of ULIGs is lower than a few \%. They are in rough agreement with the
redshift surveys of {\em IRAS} samples, in spite of some uncertainty in the data.
By construction, the IR spectra also fit the correlation of IR colours with
total IR luminosity in the Bright Galaxy Sample.  The selected scenarios are
consistent with the $\pm 1\sigma$ range for the observed CIB and they are 
used to
``disentangle'' the background into faint galaxy counts. This CIB is the first
``post--{\em IRAS}'' observation which helps constraining the high--$z$ evolution of
galaxies in the IR/submm and the energy budget integrated along the
line--of--sight, before {\em ISO} results.  In contrast with the counts at FIR
wavelengths which are rather degenerate, the submm counts are very sensitive
to the details of the evolutionary scenarios.  Thus our study illustrates the
importance of the submm range to constrain the evolution of the energy budget
of galaxies. As emphasized by Guiderdoni {\it et al.} (1997), our
understanding of the formation and evolution of galaxies is so far
entirely based on
surveys in the UV/visible window. We are missing the part of star/galaxy
formation hidden by dust. We need to open a second window in the IR/submm,
which is being explored by {\em ISO} and SCUBA, and will be extensively studied
by forthcoming telescopes and satellites such as {\em SIRTF}, SOFIA, {\em FIRST} and 
{\em PLANCK}.

There are clearly many approximations and shortcomings in our approach, which
could be listed as a research programme for forthcoming papers.  (i) We need
to implement the astrophysical processes in a code which explicitly follows
the merging history trees of haloes and galaxies.  This code would have to
include the various aspects of interaction between galaxies, and to explain
why the interaction rate strongly declined since $z=1$. It would also have to
include the effect of interactions on the SFR. (ii) The modelling of the IR
emission could take into account the evolution of dust properties with 
chemical abundances, as well as more sophisticated transfer.  (iii) The global
energy budget of galaxies has to be addressed by fitting the faint galaxy
counts in the UV/visible and $K$, as well as the forthcoming NIR counts from
{\em ISO}, in particular at 15 $\mu$m.  First results are already given here, and
suggest an amount of evolution which is consistent with our predictions. (iv)
The strong evolution of the comoving SFR density has consequences on the gas
and heavy element abundances.  The heavy elements made by the high SFRs are to
be found in the stars, the cold gas, or the hot gas of the intergalactic
medium.  If the SFR turned out to be much higher than determined, for
instance, by Madau {\it et al.} (1996), a significant fraction of the heavy
elements should be present in the IGM. The amount of elements synthetised 
in bursts would still be higher if the IMF is top--heavy, as in the simple
model of ULIGs which is used here.
According to Mushotzky and Loewenstein
(1997), the absence of evolution of the iron content of clusters at $z \sim
0.3$ is a first evidence that the global metal production is 2--5 times higher
than inferred from the UV drop--out technique.

Despite the limitations listed above, this paper provides the first
predictions of the IR luminosity function, IR/submm galaxy counts, 
redshift distributions, and
diffuse background, obtained from a semi--analytic model of galaxy formation
and evolution which takes into account the main physical processes from the
collapse of the density fluctuations to the absorption of UV and visible star
light by dust and the re--emission in the IR/submm range. The predictions of
this new model are consistent with the dynamical process of continuous galaxy
formation predicted by the hierarchical growth of structures in a SCDM
universe, and represent a significant progress with respect to previous
phenomenological models designed to make predictions in the IR/submm.

Tables giving the faint counts and redshift distributions at various
wavelengths between 15 $\mu$m and 2 mm are available by anonymous ftp
at {\bf ftp.iap.fr} in the directory /pub/from\_users/guider/fir, and on the 
Web node of the Institut d'Astrophysique de Paris at 
{\bf http://www.iap.fr/users/guider/fir.html}.

\acknowledgements{We are pleased to thank St\'ephane Charlot, Dave Clements,
Fran\c cois--Xavier D\'esert, David Elbaz, Michael Fall, Karl
Glazebrook and Jean--Loup Puget
for their comments and suggestions.}

\appendix

\section{Mass distribution of collapsed haloes}

The initial, linear perturbation is assumed to be spherical and
homogeneous. This is the so--called ``top--hat'' model, which has the friendly
property of being entirely defined by two parameters, for instance the size
$R$ and density contrast $(\delta \rho /\rho)_{z=0} \equiv \delta_0$ which are
the linearly--extrapolated values at $z=0$, or, equivalently, by the mass $M$
and collapse redshift $z_{coll}$.  If $\rho_0$ is the current mass density of
the universe, we have $M = (4\pi/3) R^3 \rho_0$.

The linearly--extrapolated density contrast, and the ratio of the radius of
maximal expansion $r_m$ to the linearly--extrapolated size $R$, can be
respectively computed as functions of the redshift of collapse:
\begin{equation}
\delta_0 = \delta_0[1+z_{coll}] , ~~~~ r_m/R = r_m/R [1+z_{coll}] .
\end{equation}
For instance, if $\Omega_0=1$, $\delta_0=\delta_c(1+z_{coll})$ with
$\delta_c=1.68$, and $r_m/R = (3/5\delta_0)$.  After collapse and violent
relaxation, a mean potential forms and the virialized halo ends as a singular,
isothermal sphere truncated at ``virial radius'' $r_V=r_m/2$.  If we define a
``circular velocity'' as $V_c \equiv (GM/r_V)^{1/2}$, the density profile of
the relaxed halo at $r \leq r_V$ is:
\begin{equation}
\rho_H(r) = {V_c^2 \over 4 \pi G r^2} ,
\end{equation}
and the mass included within radius $r$ is simply $M(<r) = Mr/r_V$.  For
instance, if $\Omega_0=1$, the virial radius (in Mpc) and the circular
velocity (in km s$^{-1}$) are:
\begin{equation} 
r_V = 0.17 ({M \over 10^{12}M_\odot})^{1/3} (1+z_{coll})^{-1}h^{-2/3} ,
\end{equation}
\begin{equation}
V_c = 160 ({M \over 10^{12}M_\odot})^{1/3} (1+z_{coll})^{1/2}h^{1/3} .
\end{equation}
We hereafter take $h \equiv H_0/(100 km s^{-1} Mpc^{-1})$.  Finally, the
virial temperature of the baryonic gas is:
\begin{equation}
T_V \equiv {1 \over 2} {\mu m_p \over k} V_c^2 = 35.9 ({V_c \over kms^{-1}})^2
~K ,
\end{equation}
$m_p$ is the mass of the proton, and the mean molecular weight $\mu =0.59$
corresponds to ionized gas with respective H and He mass fractions $X=0.75$
and $Y=0.25$.

Then we compute the mass distribution of collapsed haloes from the peaks
formalism (Bardeen {\it et al.} 1986), as in Lacey and Silk (1991) and Lacey
{\it et al.} (1993).  We start from the power spectrum of linear fluctuations
$P(k)$ which is normalized by $\sigma_8$, the variance at a {\it top--hat}
smoothing radius $R=8h^{-1}$ Mpc.  The peaks formalism uses momenta of the
power spectrum with Gaussian smoothing on radius $R_G$. The mass included
within such a radius is $M=(2\pi)^{3/2}R_G^3\rho_0$. So the relation between
the Gaussian radius $R_G$ and the top--hat radius $R$ including mass $M$ is
simply $R_G=0.64R$.  These momenta are computed from:
\begin{equation}
\sigma_m(R_G)^2 = {1 \over 8\pi^3} \int k^{2m} P(k)|W_k(kR_G)|^2 4 \pi k^2 dk ,
\end{equation}
where $W_k(y)=\exp(-y)$ is the Fourier transform of the Gaussian filtering
function:
\begin{equation}
W(x) = {1 \over (2\pi R_G^2)^{3/2}} \exp(-{|x|^2 \over 2R_G^2}) .
\end{equation}
The (comoving) number density distribution of peaks $\nu \equiv
\delta_{0G}/\sigma_0(R_G)$ in which $\delta_{0G}$ is smoothed on the Gaussian
scale $R_G$ is:
\begin{equation}
{\partial n_{pk}(R_G) \over \partial \nu}d\nu=n_{pk}(R_G) P(\nu ,R_G)d\nu ,
\end{equation}
with:
\begin{equation}
n_{pk}(R_G)={c_{\infty} \over 4 \pi^2 R_\star ^3} ,
\end{equation}
\begin{equation}
P(\nu, R_G)={1 \over c_\infty} \exp -{\nu^2 \over 2} G(\gamma ,\gamma\nu) ,
\end{equation}
such as $\int P(\nu ,R_G)d\nu=1$.  The functions $R_\star$ and $\gamma$ are
computed as:
\begin{equation}
R_\star(R_G) \equiv \sqrt 3 {\sigma_1(R_G) \over \sigma_2(R_G)} , ~~~
\gamma(R_G) \equiv {\sigma_1^2(R_G) \over \sigma_0(R_G)\sigma_2(R_G)} .
\end{equation}
The constant $c_{\infty}=0.6397$ and an analytic fit of the function $G(\gamma
,\omega)$ is given by equ. (4.4) and (4.5) of Bardeen {\it et al.} (1986).  If
$P(k) \propto k^n$, it is easy to check that $\gamma(R_G)$ does not depend on
$R_G$ and that $R_\star(R_G) \propto R_G$. For $\Omega_0=1$, $h=0.5$, and the
CDM power spectrum with $-3 \leq n \leq 1$, we have slow variations: $0.55 <
\gamma(R_G) \leq 0.8$ and $1.05 \leq R_\star(R_G)/R_G \leq 1.5$ for $0.1 \leq
R_G \leq 100$ Mpc.

At that stage, we do not have the number density distribution of peaks per bin
of density contrast {\it and} radius. For that purpose, we have to introduce,
as in Lacey \& Silk (1991), a suggestion by Bond (1988) who interprets
equ. (A9) as the total number density of peaks $n_{pk} (>R_G)$ at scale
$>R_G$. By derivating equ. (A8), it is easy to get:
\begin{equation}
{\partial^2 n_{pk} \over \partial R_G \partial \nu} dR_G d\nu = 
-{3 d \ln R_\star \over R_G d \ln R_G} n_{pk}(R_G) P(\nu ,R_G)dR_G d\nu . 
\end{equation}
We neglect the slow variation of $\gamma(R_G)$ with $R_G$.  This is the number
density of {\it all} peaks. We now need to count the peaks over a certain
threshold $\nu_{th}$ for collapse, and we get the formation rate of haloes:
\begin{equation}
{\partial^2 n_{for} \over \partial R_G \partial \nu_{th}} dR_G d\nu_{th}=
{\partial \over \partial \nu_{th}}(\int_{\nu_{th}}^{\infty} {\partial^2 n_{pk}
\over \partial R_G \partial \nu}d\nu) dR_G d\nu_{th} .
\end{equation}
After a change of variables, we get the halo formation rate, since $\nu_{th}
\equiv \delta_{0G,th}[1+z_{coll}]/\sigma_0(R_G)$:
\begin{eqnarray}
{\partial^2 n_{for} \over \partial \ln M \partial (1+z_{coll})} &=&
-n_{pk}(>R_G)
{d \ln R_\star \over d \ln R_G} P(\nu_{th},R_G) \nonumber \\ 
 & & \times {1 \over
\sigma_0(R_G)} {d \delta_{0G,th}[1+z_{coll}] \over d (1+z_{coll})} .
\end{eqnarray}
In order to link the Gaussian smoothing to the top--hat formalism, we take
$R_G=0.64R$ and $\delta_{0G,th}\equiv \delta_0 \sigma_0(R)/\sigma_0(R_G)$.
The total number of collapsed peaks at redshift $(1+z)$ is obtained by
integrating equ. (A14) on all redshifts larger than $(1+z)$.

The number density continuously increases with time, at all masses, and
contrarily to what is found from the Press--Schechter formalism.  The peaks
formalism follows the collapse of {\it all} peaks, and counts high peaks as
well as the broader, shallower peaks in which these high peaks are
included. This so--called ``cloud--in--cloud'' problem results in a possible
overestimate of the total number of galaxies. As a matter of fact, the total
number density of peaks at the low--mass end of the mass function varies as
$M^{-2}$, and the total mass density grows logarithmically with the lower mass
cut--off. Anyhow, since other astrophysical phenomena will act to ``suppress''
galaxy formation in low--mass haloes, we find that this slow divergence is not
too serious a problem.

Finally, we do not explicitly include the merging of galaxies in merging
haloes.  In our crude modelling, new galaxies form at each generation of peak
collapse.  This model could be fitted to the description of ``active'' objects
such as luminous galaxies seen in the IR. Such a simplifying assumption is very
similar to that of Haehnelt and Rees (1993) who modelled QSO formation in
each generation of halo formation. The issue of merging should be explicitly
addressed by making Monte--Carlo realizations of the halo merging history tree
and monitoring the merging of galaxies in merged haloes, for instance with the
dynamical friction time scale (Kauffmann {\it et al} 1993; Cole {\it et
al.} 1994; and following works). Such studies show that the galaxy merging
rate is relatively low, and that the resulting mass and luminosity functions
do not differ significantly from those computed with cruder assumptions. More
precisely, galaxy merging seems to be sufficient to make the 10 \% fraction of
giant galaxies which are elliptical, but does not produce a significant change
in the slope of the luminosity function at faint magnitudes.  The
implementation of the astrophysics hereafter described into this type of code
is clearly one of the next steps of our programme. Nevertheless, it is
worthwhile to note that only merging following the slow process of dynamical
friction has been currently modelled. Two--body encounters and tidal
interactions, which trigger starburst activity as it clearly appears 
from observational evidence, are not modelled yet in the current Monte--Carlo 
codes.


\begin{thebibliography}{}

\bibitem[]{} Abraham, R.G., Tanvir, N.R., Santiago, B.X., Ellis, R.S.,
Glazebrook, K., van den Bergh, S., 1996, MNRAS, 279, L47

\bibitem[]{} Andreani, P., Franceschini, A. 1992, A \& A, 260, 89

\bibitem[]{} Andreani, P., Franceschini, A., 1996, MNRAS, 283, 85

\bibitem[]{} Ashby, M.L.N., Hacking, P.B., Houck, J.R., Soifer, B.T.,
Weisstein, E.W., 1996, ApJ, 456, 428

\bibitem[]{} Bardeen, J.M., Bond, J.R., Kaiser, N., Szalay, A.S.  1986, Ap.J.,
304, 15

\bibitem[]{} Barnes, J., Efstathiou, G. 1987, ApJ, 319, 575

\bibitem[]{} Baugh, C.M., Cole, S., Frenk, C.S., 1996a, MNRAS, 283, L15

\bibitem[]{} Baugh, C.M., Cole, S., Frenk, C.S., 1996b, MNRAS, 283, 1361

\bibitem[]{} Baugh, C.M., Cole, S., Frenk, C.S., Lacey, C.G., 1997,
astro-ph/9703111

\bibitem[]{} Beichman, C.A., Helou, G. 1991, ApJL, 370, L1

\bibitem[]{} Bertin, E., Dennefeld, M., Moshir, M. 1997, A\&A, 323, 685
 
\bibitem[]{} Blain, A.W., Longair, M.S., 1993a, MNRAS, 264, 509

\bibitem[]{} Blain, A.W., Longair, M.S., 1993b, MNRAS, 265, L21

\bibitem[]{} Blanchard, A., Valls--Gabaud, D., Mamon, G., 1992, A\&A, 264, 365

\bibitem[]{} Bond, J.R., Cole, S., Efstathiou, G., Kaiser, N., 1991.  ApJ,
              379, 440

\bibitem[]{} Bond, J.R., 1988. in {\it The Early Universe}, Unruh \& Semenoff
      (eds.), p. 322

\bibitem[]{} Bower, R.G., 1991., MNRAS, 248, 332

\bibitem[]{} Briggs, F.H., Rao, S., 1993., ApJ 417, 494

\bibitem[]{} Burkey, J.M., Keel, W.C., Windhorst, R.A., Franklin, B.E., 1994,
ApJ, 429, L13

\bibitem[]{} Carlberg, R.G., 1990, ApJ, 359, L1

\bibitem[]{} Carlberg, R.G., Pritchet, C.J., Infante, L., 1994, ApJ, 435, 540

\bibitem[]{} Carico, D.P., Keene, J., Soifer, B.T., Neugebauer, G., 1992.,
   PASP, 104, 1086

\bibitem[]{} Chini, R., Kreysa, E., Kr\" ugel, E., Mezger, P.G., 1986, A\&A,
    166, L8

\bibitem[]{} Charbonnel, C., Meynet, G., Maeder, A., Schaerer, D., 1996, 
A\&ASS, 115, 339

\bibitem[]{} Chini, R., Kr\" ugel, E., 1993, A\&A, 279, 385

\bibitem[]{} Clements, D.L., Andreani, P., Chase, S.T., 1993., MNRAS, 261, 299

\bibitem[]{} Clements, D.L., Sutherland, W.J., Saunders, W., Efstathiou, G.P.,
McMahon R.G., Maddox, S., Rowan--Robinson, M., 1996a, MNRAS, 279, 459

\bibitem[]{} Clements, D.L., Sutherland, W.J., McMahon R.G., Saunders, W.,
1996b, MNRAS, 279, 477

\bibitem[]{} Cole, S., 1991., ApJ, 367, 45

\bibitem[]{} Cole, S., Arag\'on--Salamanca, A., Frenk, C.S., Navarro, J.F.,
         Zepf, S.E. 1994., MNRAS, 271, 781

\bibitem[]{} Connolly, A.J., Szalay, A.S., Dickinson, M., SubbaRao, M.U.,
Brunner, R.J., 1997, astro-ph/9706255

\bibitem[]{} Cowie, L.L., Songaila, A., Hu, E.M., Cohen, J.D., 1996, AJ, 112,
839

\bibitem[]{} Dekel, A., Silk, J., 1986, ApJ, 303, 39

\bibitem[]{} D\'esert, F.X., Boulanger, F., Puget, J.L. 1990, A\&A, 237, 215

\bibitem[]{} Disney, M., Davies, J., Philipps, S., 1989, MNRAS, 239, 939

\bibitem[]{} Draine, B.T., Lee, H.M., 1984, ApJ, 285, 89

\bibitem[]{} Dwek, E., V\'arosi, F., 1996, in {\it Unveiling the Cosmic
Infrared Background}, E. Dwek (ed), AIP Conference Proceedings 348

\bibitem[]{} Eales, S.A., Wynn--Williams, C.G., Duncan, W.D., 1989, ApJ, 339,
    859

\bibitem[]{} Eales, S.A., Edmunds, M.G., 1996a, MNRAS, 280, 1167

\bibitem[]{} Eales, S.A., Edmunds, M.G., 1996b, astro-ph/9609120

\bibitem[]{} Efstathiou, G., Rees, M.J. 1988, MNRAS, 230, 5P

\bibitem[]{} Efstathiou, G., Frenk, C.S., White, S.D.M., Davis, M.  1988,
MNRAS, 235, 715

\bibitem[]{} Efstathiou, G., 1992, MNRAS, 256, 43P

\bibitem[]{} Eisenhardt, P., Armus, L., Hogg, D.W., Soifer, B.T., Neugebauer,
G., Werner, M.W.,,1996, ApJ, 461, 72

\bibitem[]{} Ellis, R.S., Colless, M., Broadhurst, T.J., Heyl, J.S.,
Glazebrook, K., 1996, MNRAS, 280, 235

\bibitem[]{} Evrard, A.E., 1989, ApJ, 341, 26

\bibitem[]{} Fall, S.M., Efstathiou, G., 1980, MNRAS, 193, 189

\bibitem[]{} Fall, S.M., Charlot, S., Pei, Y.C., 1996, ApJ, 464, L43

\bibitem[]{} Franceschini, A., Toffolatti, L., Mazzei, P., Danese, L., \& De
Zotti, G., 1991, ApJSS, 89, 285

\bibitem[]{} Franceschini, A., Mazzei, P., De Zotti, G., Danese, L.  1994,
      ApJ, 427, 140

\bibitem[]{} Franceschini, A., Andreani, P. 1995, ApJ, 440, L5

\bibitem[]{} Gautier, T. N., III, Boulanger, F., Perault, M., Puget, J. L.,
      1992, AJ, 103, 1313

\bibitem[]{} Griffiths, R.F., {\it et al.}, 1994, ApJ, 437, 67
 
\bibitem[]{} Guiderdoni, B., 1987, A\&A, 172, 27

\bibitem[]{} Guiderdoni, B., Rocca--Volmerange, B. 1987, A\&A, 186, 1

\bibitem[]{} Guiderdoni, B., Rocca--Volmerange, B. 1988, A\&ASS, 74, 185

\bibitem[]{} Guiderdoni, B., Hivon, E., Bouchet, F.R., Maffei, B., \& Gispert,
R. 1996, in {\it Unveiling the Cosmic Infrared Background}, E. Dwek (ed), AIP
Conference Proceedings 348

\bibitem[]{} Guiderdoni, B., Bouchet, F.R., Puget, J.L., Lagache, G., Hivon,
E., 1997, {\it submitted}

\bibitem[]{} Hacking, P., Houck, J.R., 1987, ApJSS, 63, 311

\bibitem[]{} Hacking, P.B., Soifer, B.T. 1991, ApJL, 367, L49

\bibitem[]{} Haehnelt, M.G., Rees, M.J., 1993, MNRAS, 263, 168

\bibitem[]{} Hauser, M.G. 1995, in {\it Proceedings of the IAU Symp. n$^o$
    168, Examining the Big Bang and Diffuse Background Radiation}, The Hague,
    August 1994

\bibitem[]{} Helou, G., Soifer, B.T., Rowan--Robinson, M., 1985, ApJ, 298, L7

\bibitem[]{} Heyl, J.S., Cole, S., Frenk, C.S., Navarro, J.F., 1995, MNRAS,
     274, 755

\bibitem[]{} Hivon, E., Guiderdoni, B., Bouchet, F., 1997, {\it in preparation}

\bibitem[]{} Hughes, D.H., Dunlop, J.S., Rawlings, S., 1997, astro-ph/9705094

\bibitem[]{} Kauffmann, G.A.M., 1995, MNRAS, 274, 161

\bibitem[]{} Kauffmann, G.A.M., 1996, MNRAS, 281, 487

\bibitem[]{} Kauffmann, G.A.M., White, S.D.M., 1993, MNRAS, 261, 921

\bibitem[]{} Kauffmann, G.A.M., White, S.D.M., Guiderdoni, B., 1993, MNRAS,
               264, 201

\bibitem[]{} Kauffmann, G.A.M., Guiderdoni, B., White, S.D.M., 1994, MNRAS,
               267, 981

\bibitem[]{} Kauffmann, G.A.M., Charlot, S., White, S.D.M., 1996, MNRAS, 283,
               L117

\bibitem[]{} Kennicutt, R.C., 1989, ApJ, 344, 685

\bibitem[]{} Kennicutt, R.C., 1997, in {\it Starbursts: Triggers, Nature and
Evolution}, B. Guiderdoni and A. Kembhavi (eds.), Editions de Physique
/Springer--Verlag

\bibitem[]{} Kennicutt, R.C., Tamblyn, P., Congdon, C.W., 1994, ApJ, 435, 22

\bibitem[]{} Lacey, C., Silk, J., 1991, ApJ, 381, 14

\bibitem[]{} Lacey, C., Guiderdoni, B., Rocca--Volmerange, B., Silk, J., 1993,
              ApJ, 402, 15

\bibitem[]{} Lanzetta, K.M., Wolfe, A.M., Turnshek, D.A., 1995, ApJ, 440, 435

\bibitem[]{} Lilly, S.J., Tresse, L., Hammer, F., Crampton, D., Le F\`evre,
O., 1995, ApJ, 455, 108

\bibitem[]{} Lilly, S.J., Le F\`evre, O., Hammer, F., Crampton, D., 1996, ApJ,
460, L1

\bibitem[]{} Lonsdale, C.J., Hacking, P.B., Conrow, T.P., Rowan--Robinson, M.,
1990, ApJ, 358, 60

\bibitem[]{} Lobo, C., Guiderdoni, B., 1997, {\it in preparation}

\bibitem[]{} Lonsdale, C.J. 1996, in {\it Unveiling the Cosmic Infrared
Background}, E. Dwek (ed.), AIP Conference Proceedings 348

\bibitem[]{} Loveday, J., Peterson, B.A., Efstathiou, G., \& Maddox,
              S.J. 1992. ApJ, {\bf 390}, 79

\bibitem[]{} Madau, P., Ferguson, H.C., Dickinson, M.E., Giavalisco, M.,
Steidel, C.C., Fruchter, A., 1996, MNRAS, 283, 1388

\bibitem[]{} Maffei, B. 1994, PhD Dissertation, Universit\'e Paris VII

\bibitem[]{} Marzke, R.O., Huchra, J.P., Geller, M.J., 1994, ApJ, 428, 43

\bibitem[]{} Mathis, J.S., Mezger, P.G., Panagia, N., 1983, A\&A, 128, 212

\bibitem[]{} McGaugh, S.S. 1994, Nat., 367, 538

\bibitem[]{} Mushotzky, R.F., Loewenstein, M., 1997, astro-ph/9702149

\bibitem[]{} Natarajan, P., Pettini, M., 1997, astro-ph/9709014

\bibitem[]{} Natta, A., Panagia, N., 1984, ApJ, 287, 228

\bibitem[]{} Navarro, J.F., White, S.D.M., 1993, MNRAS, 265, 271

\bibitem[]{} Oliver, S.J., Rowan--Robinson, M., Saunders, W., 1992, MNRAS,
256, 15P

\bibitem[]{} Oliver, S.J., Goldschmidt, P., Franceschini, A., Serjeant,
S.B.G., Efstathiou, A.N., {\it et al.}, 1997, MNRAS, 847, 1
 
\bibitem[]{} Pearson, C., Rowan--Robinson, M., 1996, MNRAS, 283, 174

\bibitem[]{} Pei, Y.C., Fall, S.M., 1995, ApJ, 454, 69

\bibitem[]{} Pettini, M., Steidel, C.S., Dickinson, M., Kellogg, M., 
Giavalisco, M., Adelberger, K.L., 1997, astro-ph/9707200

\bibitem[]{} Puget, J.L., Abergel, A., Boulanger, F., Bernard, J.P., Burton,
W.B., {\it et al.}, 1996, A\&A, 308, L5

\bibitem[]{} Reach, W.T., {\it et al.} 1995, astro-ph 9504056

\bibitem[]{} Rice, W., Lonsdale, C.J., Soifer, B.T., Neugebauer, G., Kopan,
E.L., Lloyd, L.A., de Jong, T., Habing, H.J., 1988, ApJSS, 68, 91

\bibitem[]{} Rigopoulou, D., Lawrence, A., Rowan--Robinson, M., 1996, MNRAS,
278, 1049

\bibitem[]{} Rocca--Volmerange, B., Guiderdoni, B., 1988, A\&ASS, 75, 93

\bibitem[]{} Rowan--Robinson, M., 1992, MNRAS, 258, 787

\bibitem[]{} Rowan--Robinson, M., Broadhurst, T., Lawrence, A., McMahon, R.G.,
Lonsdale, C., {\it et al.}, 1991a, Nature, 351, 719

\bibitem[]{} Rowan--Robinson, M., Saunders, W., Lawrence, A., Leech, K., 1991b,
MNRAS, 253, 485

\bibitem[]{} Rush, B., Malkan, M., Spinoglio, L., 1993, ApJS, 89, 1

\bibitem[]{} Sanders, D.B., Mirabel, I.F., 1996, ARAA, 34, 749

\bibitem[]{} Saunders, W., Rowan--Robinson, M., Lawrence, A., Efstathiou, G.,
Kaiser, N., Ellis, R.S., Frenk, C.S., 1990, MNRAS, 242, 318

\bibitem[]{} Sawicki, M.J., Lin, H., Yee, H.K.C., 1997, AJ, 113, 1

\bibitem[]{} Schaeffer, R., Silk, J., 1985, ApJ, 292, 319

\bibitem[]{} Schaller, G., Schaerer, D., Meynet, G., Maeder, A., 1992, A\&ASS,
               96, 269

\bibitem[]{} Smith, B.J., Kleinmann, S.G., Huchra, J.P., Low, F.J., 1987, ApJ,
    318, 161

\bibitem[]{} Soifer, B.T., Sanders, D.B., Madore, B.F., Neugebauer, G.,
Danielson, G.E., {\it et al.}, 1987, ApJ, 320, 238

\bibitem[]{} Soifer, B.T., Neugebauer, G., 1991, AJ, 101, 354

\bibitem[]{} Stark, A.A., Davidson, J.A., Harper, D.A., Pernic, R.,
Loewenstein, R., Platt, S., Engargiola, G., Casem, S., 1989, ApJ, 337, 650
  
\bibitem[]{} Steidel, C.C., Giavalisco, M., Pettini, M., Dickinson, M.,
Adelberger, K.L., 1996, ApJ, 462, L17

\bibitem[]{} Storrie--Lombardi, L.J., McMahon, R.G., Irwin, M.J., 1996, MNRAS,
283, L79

\bibitem[]{} Thornton, K., Gaudlitz, M., Janka, H.T., Steinmetz, M., 1997,
astro-ph/9706175

\bibitem[]{} Wang, B., 1991a, ApJ, 374, 456

\bibitem[]{} Wang, B., 1991b, ApJ, 374, 465

\bibitem[]{} White, S.D.M., Rees, M.J., 1978, MNRAS, 183, 341

\bibitem[]{} White, S.D.M., Frenk, C.S., 1991, ApJ, 379, 52

\bibitem[]{} Williams, R.E., {\it et al.}, 1996, AJ, 112, 1335

\bibitem[]{} Zurek, W.H., Quinn, P.J., Salmon, J.K., 1988, ApJ, 330, 519

\bibitem[]{} Zepf, S.E., Koo, D.C., 1989, ApJ, 337, 34

\end{thebibliography}
\end{document}